\documentclass[twocolumn,superscriptaddress,prl,amsmath,amstex,amssymb,citeautoscript,longbibliography]{revtex4-1}
\pdfoutput=1

\usepackage{rotating}
\usepackage{natbib}
\usepackage[english]{babel}
\usepackage{letltxmacro}
\usepackage{latexsym}
\usepackage[utf8]{inputenc}
\LetLtxMacro{\ORIGselectlanguage}{\selectlanguage}
\makeatletter
\DeclareRobustCommand{\selectlanguage}[1]{%
  \@ifundefined{alias@\string#1}
    {\ORIGselectlanguage{#1}}
    {\begingroup\edef\x{\endgroup
       \noexpand\ORIGselectlanguage{\@nameuse{alias@#1}}}\x}%
}
\newcommand{\definelanguagealias}[2]{%
  \@namedef{alias@#1}{#2}%
}
\makeatother
\definelanguagealias{en}{english}
\definelanguagealias{English}{english}
\usepackage{amsmath}
\usepackage{amsfonts}
\usepackage{amsthm}   
\usepackage{amssymb}
\usepackage{bm}
\usepackage{color}
\usepackage[percent]{overpic}
\usepackage{soul} 
\usepackage{amssymb}
\usepackage{wasysym}
\usepackage{dsfont}
\usepackage{float}
\usepackage{thm-restate}
\usepackage{mathtools}

\usepackage{hyperref}
\hypersetup{
    bookmarks=false,         
    unicode=false,          
    pdftoolbar=false,        
    pdfmenubar=true,        
    pdffitwindow=false,     
    pdfstartview={FitH},    
    pdftitle={},    
    pdfauthor={Authors},     
    pdfsubject={},   
    pdfcreator={},   
    pdfproducer={}, 
    pdfkeywords={quantum many-body scars} {Fermi-Hubbard} {constrained spin models}, 
    pdfnewwindow=true,      
    colorlinks=true,       
    linkcolor=black,          
    citecolor=blue,        
    filecolor=magenta,      
    urlcolor=blue           
}

\usepackage{cleveref}

\usepackage{graphicx}
\usepackage[colorinlistoftodos]{todonotes}
\usepackage{verbatim}

\newcommand{\ket}[1]{\mbox{$| #1 \rangle$}}

\newcommand{\HG}{\ket{{-}{+}}}
\newcommand{\HGG}{\ket{{+}{-}}}
\newcommand{\isol}{\ket{\downarrow 2 \uparrow}}
\newcommand{\isoll}{\ket{\uparrow 2 \downarrow}}
\newcommand{\Se}{S_{\mathrm{ent}}}

\usepackage{soul}
\usepackage[normalem]{ulem}

\newtheorem*{lemma*}{Lemma}

\usepackage{algorithm}
\usepackage{algpseudocode}

\newcommand*{\id}{{\normalfont\hbox{1\kern-0.15em \vrule width .8pt depth-.5pt}}}

\begin{document}

\title{A proposal for realising quantum scars in the tilted 1D Fermi-Hubbard model}

\author{Jean-Yves Desaules}
\affiliation{School of Physics and Astronomy, University of Leeds, Leeds LS2 9JT, United Kingdom}
\author{Ana Hudomal}
\affiliation{School of Physics and Astronomy, University of Leeds, Leeds LS2 9JT, United Kingdom}
\affiliation{Institute  of  Physics  Belgrade,  University  of  Belgrade,  11080  Belgrade,  Serbia}
\author{Christopher J. Turner}
\affiliation{School of Physics and Astronomy, University of Leeds, Leeds LS2 9JT, United Kingdom}
\author{Zlatko Papi\'c}
\affiliation{School of Physics and Astronomy, University of Leeds, Leeds LS2 9JT, United Kingdom}

\date{\today}

\begin{abstract}
Motivated by recent observations of ergodicity breaking due to Hilbert space fragmentation in 1D Fermi-Hubbard chains with a tilted potential [Scherg \emph{et al.}, arXiv:2010.12965], we show that the same system also hosts quantum many-body scars in a regime $U\approx \Delta \gg J$ at electronic filling factor $\nu=1$. We numerically demonstrate that the scarring phenomenology in this model is similar to other known realisations such as Rydberg atom chains, including persistent dynamical revivals and ergodicity-breaking many-body eigenstates. At the same time, we show that the mechanism of scarring in the Fermi-Hubbard model is different from other examples in the literature: the scars originate from a subgraph, representing a free spin-1 paramagnet, which is weakly connected to the rest of the Hamiltonian's adjacency graph. Our work demonstrates that correlated fermions in tilted optical lattices provide a platform for understanding the interplay of many-body scarring and other forms of ergodicity breaking, such as localisation and Hilbert space fragmentation.
\end{abstract}

\maketitle{}

{\sl Introduction.} Recently, there has been much interest in understanding how closed many-body quantum systems evolve in time when taken out of their equilibrium state. While many such systems rapidly return to their equilibrium state, in accordance with fundamental principles of quantum statistical mechanics~\cite{Gogolin2016},  much of recent work has focused on systems that fail to do so as a consequence of ergodicity breaking~\cite{Huse-rev, AbaninRMP}, either due to  the special mathematical structure known as integrability or very strong disorder which leads to (many-body) localisation. Both of these paradigms of behaviour are actively investigated in experiments on cold atoms, trapped ions and superconducting qubits~\cite{Kinoshita06, Bloch15, Monroe16,Chiaro2020}. 

The inability of non-ergodic systems to act as heat reservoirs for their smaller parts has been traditionally known to affect the entire spectrum of the system.  Recently, however, there has been a flurry of interest in \emph{weak} ergodicity breaking phenomena~\cite{Serbyn2020}. The latter refers to the emergence of  a dynamically-decoupled subspace within the many-body Hilbert space, in general without any underlying symmetry, spanned by ergodicity-breaking eigenstates.  This behaviour was first theoretically established in the Affleck-Kennedy-Lieb-Tasaki (AKLT) model~\cite{Arovas1989, BernevigEnt}, followed by the discovery of similar phenomenology in other non-integrable lattice models~\cite{Iadecola2019_2, Iadecola2019_3, Bull2019, Chattopadhyay,OnsagerScars, MoudgalyaFendley, Surace2020, Kuno2020},  models of correlated fermions and bosons~\cite{Vafek, MotrunichTowers, MarkHubbard,MoudgalyaHubbard, Moudgalya2019, bosonScars, Zhao2020}, frustrated magnets~\cite{Lee2020, McClarty2020}, topological phases of matter~\cite{NeupertScars, Wildeboer2020}, and periodically driven systems~\cite{Buca2019, BucaHubbard, Sugiura2019,Mizuta2020, Mukherjee2020}. In these examples, the ergodicity-breaking eigenstates are either explicitly embedded into a many-body spectrum via the mechanism due to Shiraishi and Mori~\cite{ShiraishiMori}, or they form a representation of an algebra~\cite{Pakrouski2020, Ren2020, Dea2020}. 

A well-known example of weak ergodicity breaking in single-particle systems is the phenomenon of quantum scars in chaotic stadium billiards~\cite{Heller84}. In this case, the particle's eigenfuctions exhibit anomalous concentration in the vicinity of an unstable periodic orbit in the classical limit $\hbar\to0$~\cite{HellerLesHouches, Polavieja1994, Wisniacki2001}, leading to observable consequences in many physical systems~\cite{Sridhar1991, Wilkinson1996, Wintgen1989, Arranz1998}.
 In recent experiments on interacting Rydberg atom arrays~\cite{Bernien2017}, weak ergodicity breaking was observed via persistent revivals following the global quench of the system, prompting the name ``quantum many-body scarring"~\cite{Turner2017,wenwei18TDVPscar,Turner2020} by analogy with stadium billiards~\cite{Choi2018,  Bull2020}. Recently, quantum many-body scarring has been shown to occur in higher dimensions~\cite{Michailidis2D,Hsieh2020,Bluvstein2020} and in the presence of certain kinds of perturbations~\cite{TurnerPRB, Khemani2018, Lin2020} including disorder~\cite{MondragonShem2020}.

On the other hand, it has also been shown that ergodicity breaking can occur due to a fracturing of the Hilbert space into dynamically disconnected components~\cite{Khemani2019_2,Pretko2019, MoudgalyaKrylov,Sala2019}. 
This typically occurs by the interplay of local interactions  with a higher-moment symmetry such as charge dipole conservation, which non-trivially intertwines spatial and internal symmetries. 
Recent work~\cite{Scherg} has demonstrated that Hilbert space fragmentation can be experimentally realised via magnetic field gradient applied to the Fermi-Hubbard (FH) model in a 1D optical lattice.
Apart from offering a new platform to investigate the link between fragmentation and the so-called Stark many-body localisation~\cite{Schulz2019, Nieuwenburg2019, Yao2020}, an immediate question presents itself: can the tilted FH model realise quantum many-body scars?

In this paper, we show that quantum many-body scars arise in the limit $U\approx \Delta\gg J$ in the tilted FH model, and that they can be detected using the quench from a specific initial state at a different filling factor from the one considered in Ref.~\onlinecite{Scherg}.
We derive an effective model for this setup, which can be mapped to a spinful generalisation of the fractional quantum Hall effect on a thin torus~\cite{Moudgalya2019}, allowing for a practical experimental realisation.
While the phenomenology of quantum many-body scars is shown to be largely similar to their realisation in Rydberg atom systems~\cite{Bernien2017}, including in particular an extensive set of eigenstates which violate the Eigenstate Thermalisation Hypothesis (ETH)~\cite{DeutschETH,SrednickiETH},  the origin of scars is different in the two systems and can be intuitively understood from a graph-theoretic viewpoint.

{\sl Large-tilt limit of the FH model.}  The 1D FH model is given by the Hamiltonian
\begin{eqnarray}\label{eq:hamfull}
\hat{H} \!  = \!  \sum_{j,\, \sigma=\uparrow ,\, \downarrow} \!  -J\hat{c}^\dagger_{j,\sigma}\hat{c}_{j+1,\sigma}^{} +{\rm h.c.}+\Delta j\hat{n}_{j,\sigma}   +  U\sum_j{\hat{n}_{j,\uparrow} \hat{n}_{j,\downarrow} }, \;\;\;\;\;
\end{eqnarray}
where $\hat{c}_{j,\sigma}^\dagger$ denotes the usual electron creation operator on site $j$ with spin projection $\sigma$, $\hat n_{j,\sigma} \equiv \hat{c}_{j,\sigma}^\dagger \hat{c}_{j,\sigma}$ , $J$ and $U$ are the hopping and on-site interaction terms, respectively. Tilt of the optical lattice is parametrised by $\Delta$, which we take to be spin-independent~\cite{Scherg}. Note that tilting has the structure of a dipole term, $\sim j \hat  n_{j}$. Below we impose open boundary conditions on the model in Eq.~(\ref{eq:hamfull}), and restrict to the electron filling factor $\nu=1$, i.e., with $N/2$ fermions with spin $\uparrow$ and $N/2$  fermions with spin $\downarrow$ on a chain of $N$ sites (assumed to be even). We also set $J=1$ for simplicity. We label the Fock states using  $\uparrow$ to denote a fermion with spin up and $\downarrow$ with spin down, while $0$ stands for an empty site and $\updownarrow$ denotes a doublon.

We focus on the regime $\Delta\approx U\gg J$. In this case the sum of the dipole moment and the number of doublons is effectively conserved. The dominant contribution to the Hamiltonian (using a Schrieffer-Wolff transformation at first order~\cite{Bravyi2011}) is then given by
\begin{eqnarray}\label{eq:hameff}
\notag	\hat{H}_{\rm eff} &=&-J\sum_{j,\sigma}  \hat{c}^{\dagger}_{j,\sigma}\hat{c}_{j+1,\sigma}\hat{n}_{j,\overline{\sigma}}(1-\hat{n}_{j+1,\overline{\sigma}}) + \mathrm{h.c.}
\\
 &+& (U-\Delta)\sum_j{\hat{n}_{j,\uparrow} \hat{n}_{j,\downarrow} }.
\end{eqnarray}
In this effective Hamiltonian, hopping to the left (which decreases the total dipole moment by 1) is only allowed if it increases the number of doublons by the same amount ($\bar{\sigma}$ denotes opossite spin from $\sigma$).

The action of the Hamiltonian (\ref{eq:hameff}) within the $\nu=1$ sector fragments the Hilbert space beyond the simple conservation of $U+\Delta$. In this work we focus on the largest connected component, which is the one containing the state with alternating $\uparrow$ and $\downarrow$ fermions. In addition to the symmetries of the full model in Eq.~(\ref{eq:hamfull}), i.e., SU(2) spin symmetry and spin reversal \cite{1D_Hubbard}, 
the Hamiltonian (\ref{eq:hameff}) projected to the largest sector has an additional symmetry related to spatial inversion and particle-hole exchange~\cite{SOM}. After resolving these symmetries, we find the level statistics parameter $\langle r \rangle$~\cite{OganesyanHuse} to be close to 0.53 for all symmetry sectors with large numbers of states ($\gtrsim 10^3$)~\cite{SOM}. From these values which coincide with the Wigner-Dyson statistics~\cite{Mehta2004}, we expect the model in Eq.~(\ref{eq:hameff}) to be chaotic. We next outline an intuitive approach for identifying many-body scars in this model. 

\begin{figure}[htb]
	\centering
	\includegraphics[width=\linewidth]{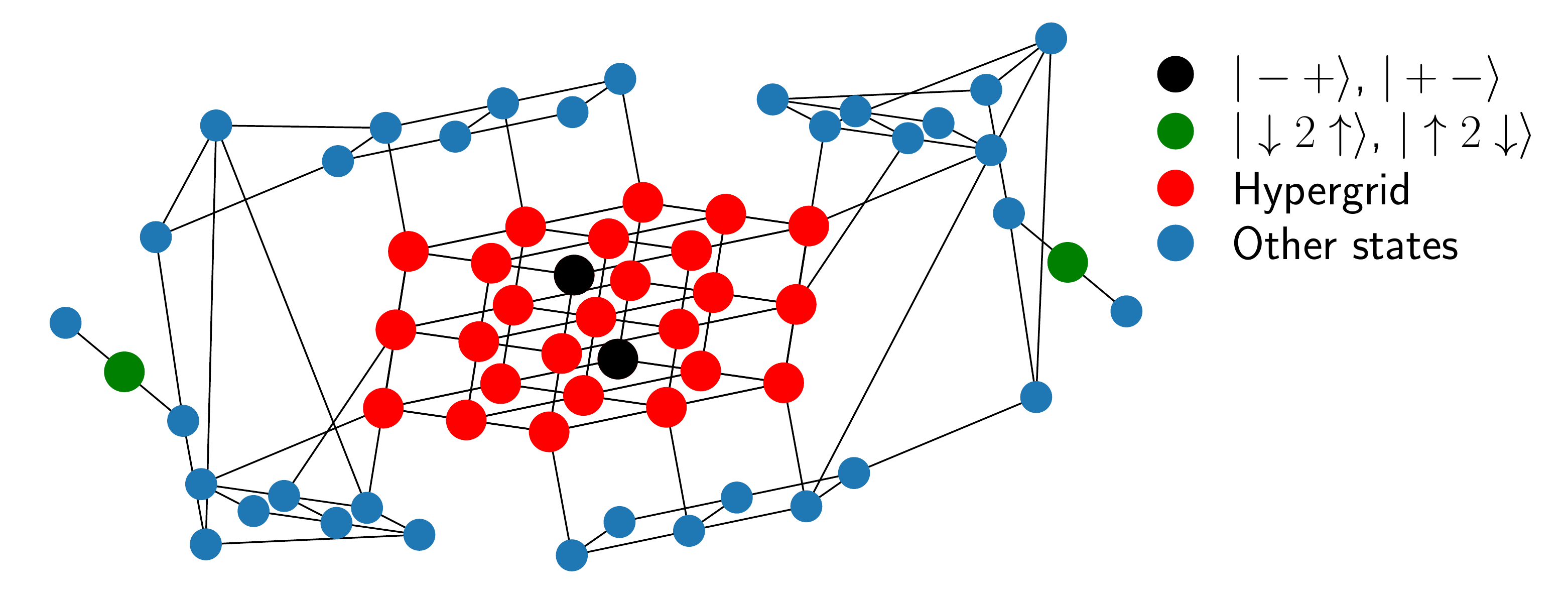}
	\caption{ Adjacency graph of the effective model in Eq.~(\ref{eq:hameff}) for $N=6$. Red vertices denote the states belonging to the hypergrid, with the black vertices corresponding to $\HG$, $\HGG$ states defined in the text. Green vertices are the isolated states $\isol$, $\isoll$ which live on the tails of the graph. For this graph, hypergrid contains 27 vertices out of 63.
	}
	\label{fig:HG_graph_6}
\end{figure}
{\sl Embedded hypergrid subgraph.} A practical diagnostic of quantum many-body scars is the existence of weakly-correlated states which undergo robust revivals under quench dynamics, while the majority of other initial states thermalise fast and do not display revivals. 
In the Rydberg-blockaded chains~\cite{Bernien2017}, the reviving N\'eel state of atoms is the densest configuration compatible with the blockade constraint, and it is an extremal vertex of the Hamiltonian adjacency graph~\cite{Turner2017}. In this graph each vertex corresponds to a basis state, and two vertices are connected by an edge if the Hamiltonian matrix element between their respective basis states is non-zero.  We next show, by examining the adjacency graph of the model in Eq.~(\ref{eq:hameff}), that we can identify a subgraph, weakly coupled to the rest of the Hilbert space, which contains the reviving initial states and leaves a strong imprint on the scarred eigenstates. This  leads to a transparent manifestation of scarring in the original Fock basis, in contrast with Rydberg atoms. In the latter case, the subspace which is weakly coupled to the rest of the Hilbert space has a much more complicated structure, leading to the wavefunction spreading across the entire adjacency graph~\cite{TurnerPRB} before refocusing onto the Néel state.

In Fig.~\ref{fig:HG_graph_6} we plot the adjacency graph of the Hamiltonian in Eq.~(\ref{eq:hameff}) for a small system. 
For the effective model in Eq.~(\ref{eq:hameff}), it is possible to gauge away the fermionic minus signs~\cite{SOM}, resulting in an unweighted, undirected graph.  As the Hamiltonian (\ref{eq:hameff}) (for $U=\Delta$) has no diagonal elements and the spectrum is symmetric around zero, all product states are effectively in the infinite temperature ensemble and are expected to thermalise quickly. As we confirm numerically below, there are two important exceptions. 

First, as highlighted in red colour in Fig.~\ref{fig:HG_graph_6}, there is a regular subgraph which has the form of the hypergrid -- a Cartesian product of line graphs (in our case, of length 3), i.e., the hypergrid is isomorphic to an adjacency graph of a free spin-1 paramagnet. This mapping can be understood by looking at the state $\ket{\downarrow \uparrow\downarrow \uparrow\downarrow \uparrow\ldots}$. Each cell of two sites can take the values ${-}\coloneqq  \downarrow \uparrow $, $2\coloneqq \updownarrow 0 $ or ${+}\coloneqq \uparrow \downarrow$, leading to a three level system. Note that the configuration $0\updownarrow$ is omitted, as doublons can only be formed by hopping to the left. On the other hand, hopping between two neighbouring cell will break this mapping and take the system out of the hypergrid subgraph.
Inside the hypergrid, we identify two states for which the cell alternates between $-$ and $+$. These are the state $\HG\coloneqq \ket{{-}{+}{-}{+}\ldots}=\ket{\downarrow \uparrow \uparrow \downarrow\downarrow \uparrow \uparrow \downarrow\ldots}$ and its spin-inverted partner, $\HGG=\ket{{+}{-}{+}{-}\ldots}\coloneqq \ket{ \uparrow\downarrow \downarrow \uparrow \uparrow\downarrow \downarrow \uparrow\ldots}$.  The states $\HG$ and $\HGG$ for $N=6$ are shown in black colour in Fig.~\ref{fig:HG_graph_6}. These two states are the only corners of the hypergrid (state with only $+$ and $-$ cells) with no edges going out of it. As we show below, either of these states shows persistent oscillations in quench dynamics, undergoing robust state transfer to their spin-inverted counterpart. While other corners of the hypergrid also show revivals, they are much smaller in amplitude and decay faster due to the leakage out of this substructure.
The second example of a reviving state is $\isol\coloneqq \ket{\downarrow \downarrow \ldots \downarrow \updownarrow 0 \uparrow \uparrow \ldots \uparrow}$ (and its spin-reversed partner $\isoll$), which is situated on a tail-like structure of length 3 (independent of system size) with minimal connectivity to the rest of the Hilbert space (green points in Fig ~\ref{fig:HG_graph_6}).  Similar tail-like structures occur in constrained spin models~\cite{Pancotti2020}.
 
{\sl Many-body scarred dynamics and eigenstates.}  Having identified candidate states for revivals, we now scrutinise their quench dynamics using large-scale exact diagonalisation simulations of the effective model in Eq.~(\ref{eq:hameff}). Making use of various symmetries present in the model, we have been able to exactly simulate dynamics for up to $N=22$ electrons. For convenience, the simulations were performed in the spin representation of the model~\cite{SOM}.

\begin{figure}[t]
	\centering
	\includegraphics[width=\linewidth]{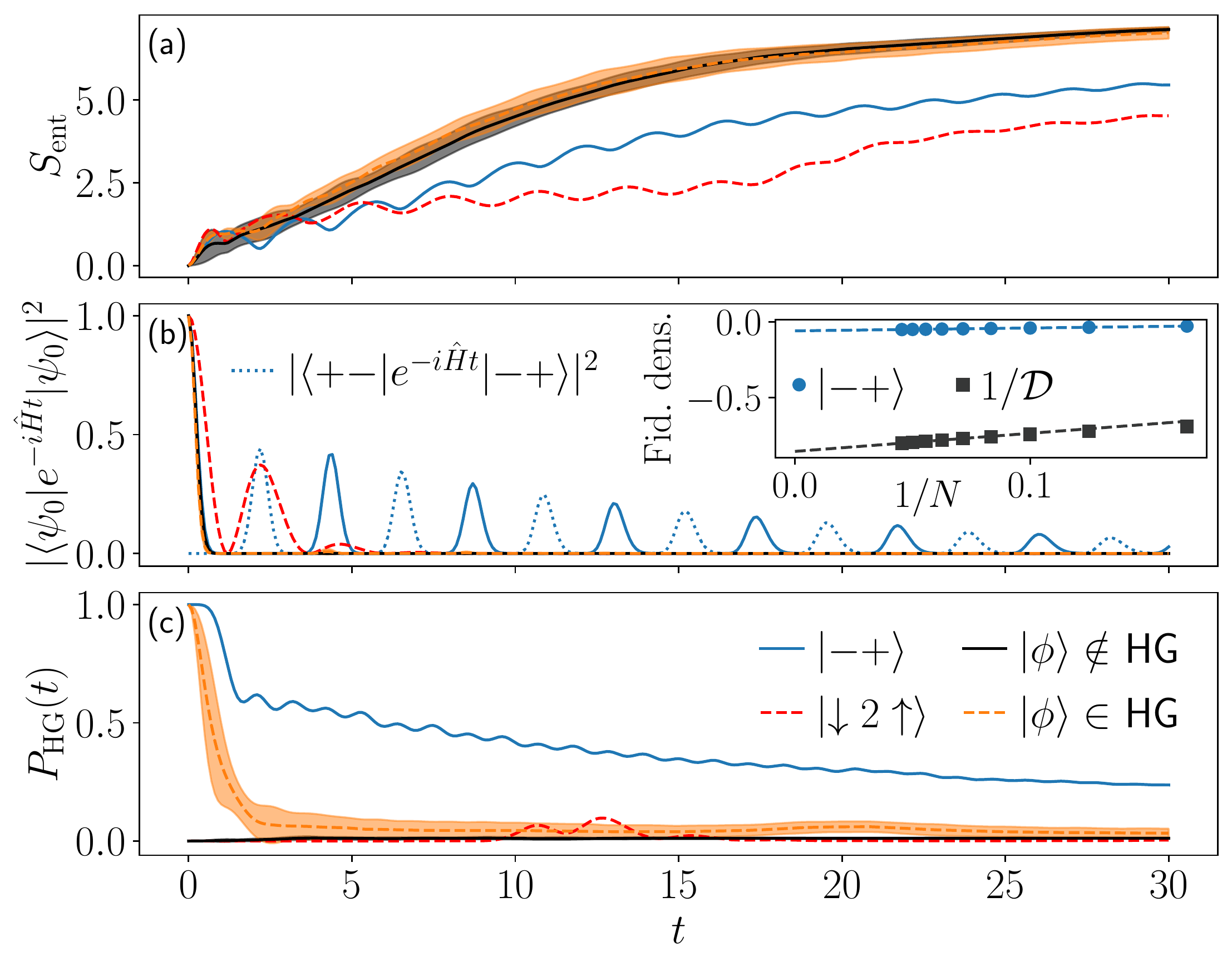}
	\caption{ Dynamics in the effective model (\ref{eq:hameff}) for $N=18$ for $\HG$,  $\isol$ and randomly chosen initial states.
		(a) Entanglement entropy $\Se$ for an equal bipartition of the system. Entropy grows linearly in time for all states, consistent with thermalising dynamics, but it shows oscillations due to many-body scarring.
				$\ket{\phi}\notin$ HG and $\ket{\phi}\in$ HG denote the average over 10 random product states outside or within the hypergrid, respectively, and the shading represents standard deviation.
		(b) Fidelity dynamics for the same initial states as in (a). Inset  shows the finite-size scaling of the fidelity density $\frac{1}{N}\ln |\langle\psi_0|e^{-i \hat H t}|\psi_0\rangle|^2$ at the first revival for  $\HG$ state, demonstrating a value much higher than $\frac{1}{N}\ln(1/\mathcal{D})$ (with $\mathcal{D}$ the dimension of the Hilbert space), expected for a random state. Blue dotted line shows the amplitude of state transfer between $\HG$ and  $\HGG$ states. 
		(c) Probability to remain within the hypergrid over time is much higher for $\HG$ than other states.
	}
	\label{fig:HG_eff_dynamics}
\end{figure}
Fig.~\ref{fig:HG_eff_dynamics}(a) shows the time dependence of the entanglement entropy $\Se(t)$ when the system is quenched from various initial product states, such as $\ket{-+}$, $\ket{\downarrow 2 \uparrow}$ and a few randomly-chosen product states.   $S_{\mathrm{ent}}$ is defined as the von Neumann entropy of the reduced density matrix for one half of the chain.  
In all cases, entropy grows linearly in time,  consistent with thermalisation of the system. However, the coefficient of linear growth is visibly different for $\ket{-+}$ and $\ket{\downarrow 2 \uparrow}$ states, and it is smaller than that of random states, indicating non-ergodic dynamics. The long-time value of the entropy is also different for the $\HG$ state \cite{SOM}, hinting that the wavefunction is still not completely spread into the whole Hilbert space. The hallmark of many-body scars is the oscillations superposed on top of the linear growth, as seen in the scarred dynamics in Rydberg atom chains~\cite{Turner2017}. Rapid growth of entropy at short times is a consequence of the bipartition being located in the middle of an two-site effective cell.

Entropy oscillations mirror those of the wavefunction return probability,$|\langle \psi_0|e^{-i \hat H t}|\psi_0 \rangle|^2$, in Fig.~\ref{fig:HG_eff_dynamics}(b). For the isolated state $\isol$, only a single revival is clearly visible as the return probability decays rapidly once the wavefunction leaks out of the tail of the graph. Because of the low connectivity of the tail, the first revival is still visible on the scale of Fig.~\ref{fig:HG_eff_dynamics}(b). The revival time can be accurately estimated by assuming the tail is completely disconnected, leading to the period $\pi/\sqrt{2}$.  In contrast, the state $\HG$ displays several revivals with the sizable weight of the wavefunction $\sim 40\%$ returning to its initial value.  The fidelity density, $\frac{1}{N}\ln |\langle\psi_0|e^{-i \hat H t}|\psi_0\rangle|^2$, shown in the inset, converges as $1/N$ to a value of $-0.058$. In contrast, the inverse Hilbert space dimension, $\mathcal{D}^{-1}$, expected for a random state leads to a fidelity density of $-0.855$ -- an order of magnitude higher.
The scarred dynamics in this case can be visualised as the state bouncing within the hypergrid between $\HG$ and its partner $\HGG$, illustrated by the dotted line in Fig.~\ref{fig:HG_eff_dynamics}(b). From the hypergrid analysis, we expect the revival period to be $\sqrt{2} \pi$, coming from the $2\pi$ period of free precession and the spin-1 matrix elements $\sqrt{2}$. This prediction closely matches the revival period observed in Fig.~\ref{fig:HG_eff_dynamics}(b). 

The importance of the hypergrid  for scarred dynamics is illustrated in Fig.~\ref{fig:HG_eff_dynamics}(c) which plots the probability to remain in the hypergrid, $P_{\mathrm{HG}}(t) = \langle \psi_0 | e^{i \hat{H} t } \hat{P}_\mathrm{HG} e^{-i \hat H t}| \psi_0\rangle$, where $\hat{P}_\mathrm{HG}$ is the projector onto the subspace spanned by product states belonging to the hypergrid. For the initial state $\HG$, we observe that the wavefunction remains concentrated inside the hypergrid, even at late times. This is in stark contrast with the PXP model which describes a chain of Rydberg atoms in the blockade regime~\cite{TurnerPRB}, where the wavefunction spreads across the entire graph by the time it undergoes the first revival. 
Furthermore, even at the first revival peak the fidelity is lower than $P_{\mathrm{HG}}$. This shows that the wavefunction does not exactly return to itself but gets more spread even within the hypergrid. 
Finally, for this initial state after a long time $P_{\mathrm{HG}}$ converges to a non-zero value which is higher than expected from the relative size of the hypergrid in the Hilbert space~\cite{SOM}, hinting that the subgraph could have additional structure that prevents states from leaking out. The fact that this long-time value is much lower for random states in the hypergrid than for $\HG$ confirms that this is not simply due to low connectivity between the hypergrid and the rest of the Hilbert space, but that the special eigenstates indeed play an important role.

\begin{figure}[t]
	\centering
	\includegraphics[width=\linewidth]{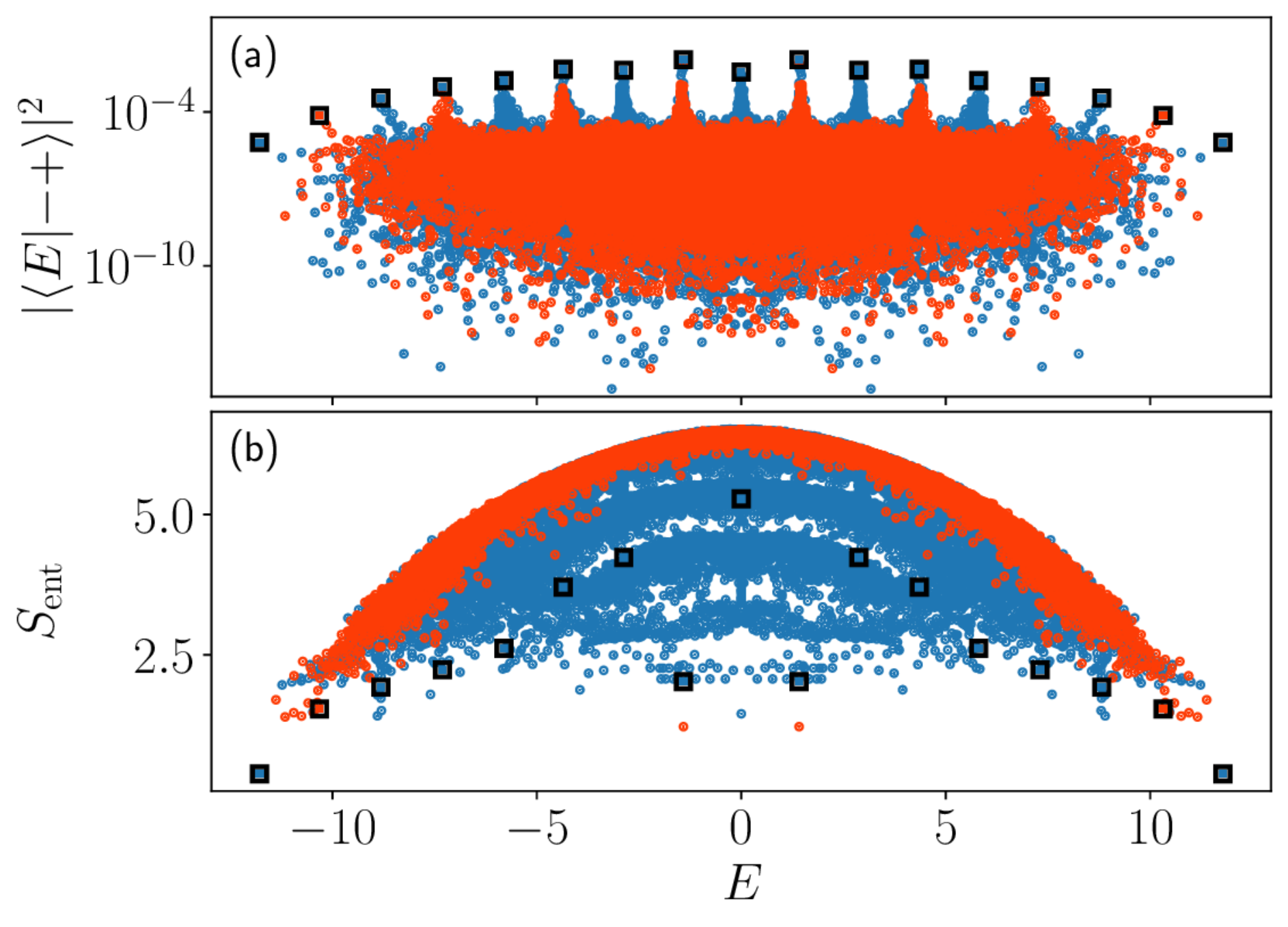}
	\caption{ Eigenstate properties of the effective model (\ref{eq:hameff}).  (a) Overlap of eigenstates with the $\HG$ state as a function of their energy $E$. (b) Entanglement entropy $\Se$ of the eigenstates. Data is for system size $N=16$. Red dots correspond to eigenstates with total spin $S=1$, while the blue ones ones mark all other spin values. The squares indicate the eigenstates sitting at the top of each tower of states. These towers have an energy separation of approximately $\sqrt{2}$, as expected for the spin-1 hypergrid.
	}
	\label{fig:eff_E_S}
\end{figure}
Properties of eigenstates of the model (\ref{eq:hameff}) are summarised in Fig.~\ref{fig:eff_E_S}. The projection of eigenstates onto  $\HG$ state, shown in panel (a), displays prominent tower structures reminiscent of other scarred models~\cite{Turner2017, Bull2019}. The existence of towers implies that eigenstates tend to concentrate around certain energies in the spectrum, causing an ETH violation. The separation between the towers is approximately $\Delta E \approx \sqrt{2}$, as expected from the embedded hypergrid. 
Note that the eigenstates have been classified according to the conserved total value of spin $S$; in contrast, $\HG$ state is not an eigenstate of $\mathbf{S}^2$. One can show that for this state,  $\langle \mathbf{S}^2 \rangle= N/2$, thus $\HG$  is predominantly supported by $S=1$ and $S=2$ eigenstates at the given system size. The $S=1$ eigenstates are indicated by red points in  Fig.~\ref{fig:eff_E_S}. 

Similar violation of the ETH can be seen in the large spread in entanglement entropy of eigenstates in Fig.~\ref{fig:eff_E_S}(b), showing that eigenstates of similar energy have very different amounts of entanglement. Part of this spreading, however, can be attributed to the eigenstates belonging to different spin sectors $S$, giving rise to multiple bands that do not fully overlap at the system size shown in Fig.~\ref{fig:eff_E_S}(b)~\cite{SOM}. The distribution of entropy in $S=1$ sector [red points in Fig.~\ref{fig:eff_E_S}(b)] is relatively narrow apart from two ``outliers" shown at energy $E\approx \pm \sqrt{2}$, which sit at the top of the tower for their sector in  Fig.~\ref{fig:eff_E_S}(a). The states at the top of each tower are indicated by squares, but unlike the PXP model~\cite{TurnerPRB} these states are not well separated from other states in the same tower. 

\begin{figure}[b]
	\centering
	\includegraphics[width=0.49\linewidth]{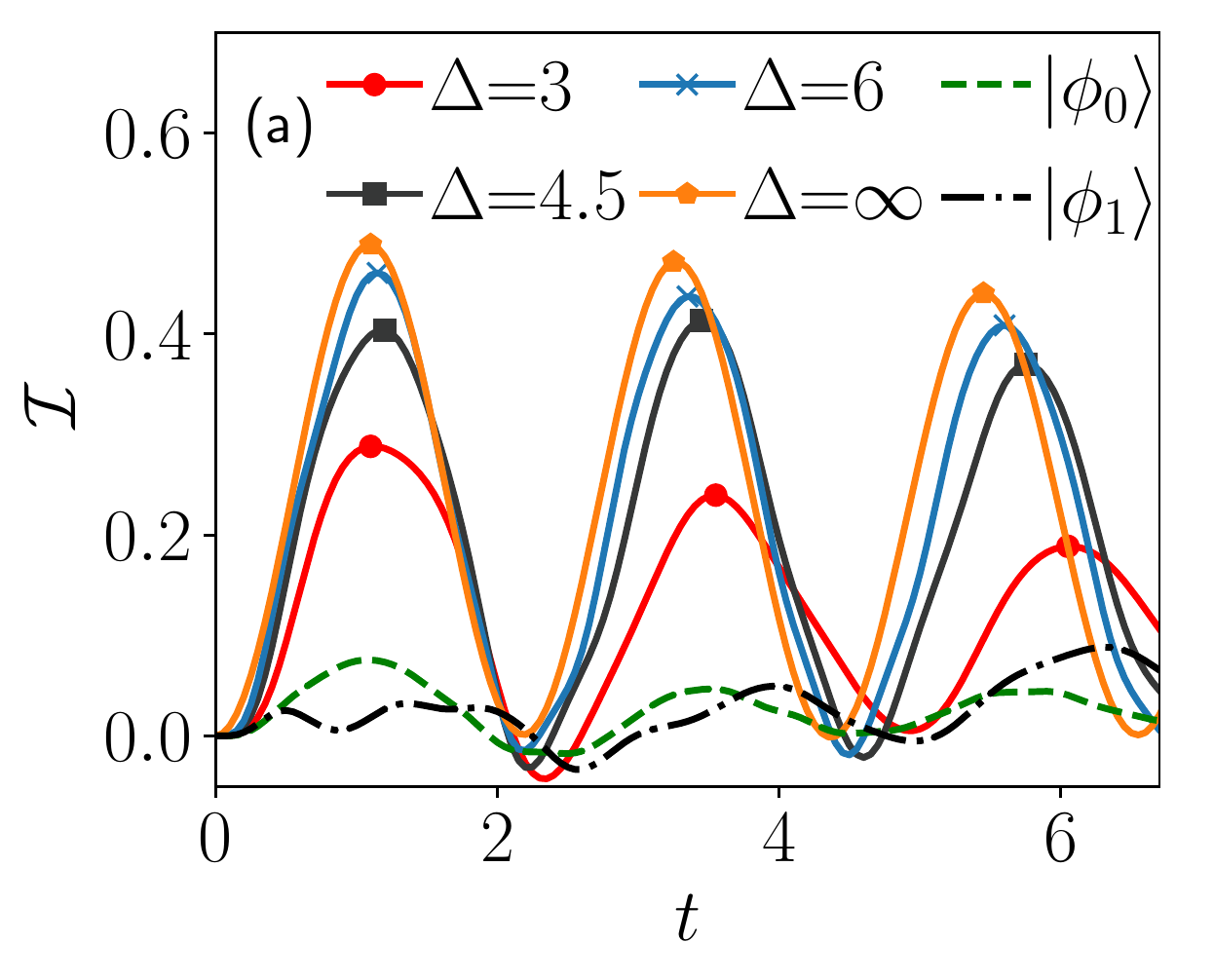}
	\includegraphics[width=0.49\linewidth]{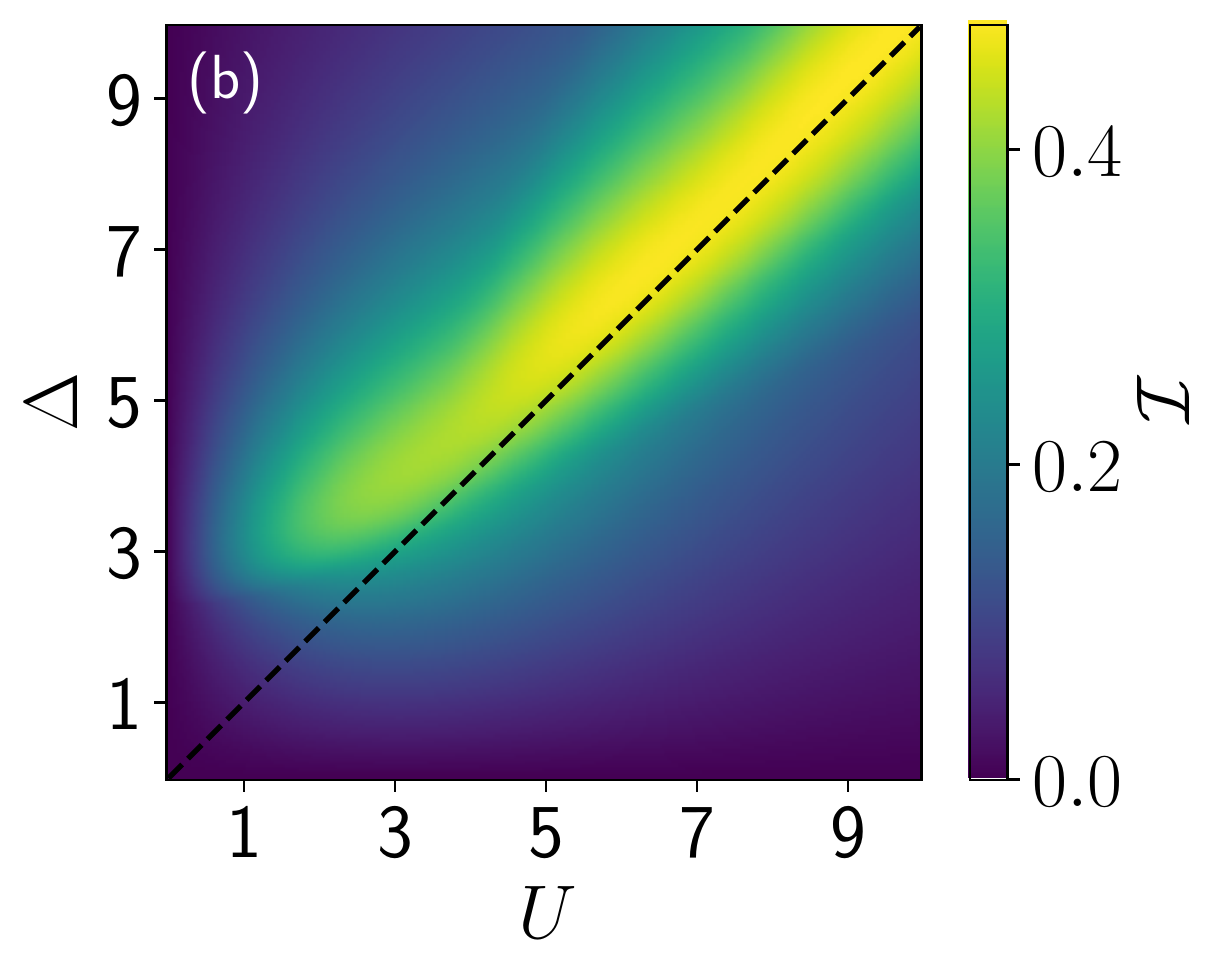}
	
	\caption{(a) Occupation imbalance in the full model (in Eq.~(\ref{eq:hamfull})) with $N{=}12$ for various values of $U{=}\Delta$ for the initial state $\HG$, and for $U{=}\Delta{=}6$ for the initial states $\ket{\phi_0}=\ket{\uparrow \downarrow\uparrow \downarrow \cdots \uparrow \downarrow }$ (within the hypergrid) and $\ket{\phi_1}=\ket{\downarrow\updownarrow 0 \uparrow \updownarrow 0 \updownarrow 0 \downarrow \updownarrow 0 \uparrow}$ (outside of it). (b) $U-\Delta$ phase diagram showing the scarring regime near the diagonal (dashed line). The colour scale represents the value of the first peak of the imbalance for $N{=}12$.
	}
	\label{fig:exp_data}
\end{figure}
{\sl Experimental implications.} The effective model studied above is exact for $U=\Delta\rightarrow \infty$. For experimental realisations, it is important to ascertain that the same physics persists for accessible values of $U$, $\Delta$ and that it can be detected using local measurements. We demonstrate this in Fig.~\ref{fig:exp_data} for the full model in Eq.~(\ref{eq:hamfull}) focusing on the regime $U, \Delta < 10$. Panel (a) shows the dynamics of imbalance on the even/odd sublattices, $\mathcal{I}=(N_{\rm o}-N_{\rm e})/(N_{\rm o}+N_{\rm e})$, where $N_{\rm e/o}$ is the total number of fermions on the even or odd sites. 
 The imbalance is bounded between -1 and 1.  We see robust oscillations in $\mathcal{I}$ with the frequency matching  half the wavefunction revival frequency in Fig.~\ref{fig:HG_eff_dynamics}(b). The amplitude of the imbalance revival remains close to the infinite-limit value for $U=\Delta\gtrsim 6$. 
As further evidence that the hypergrid is the cause of non-ergodicity, we devised a local perturbation which effectively disconnects the hypergrid, leading to the improvement of revivals in the full model ~\cite{SOM}.

{\sl Conclusions and discussion.} We have proposed an experimental realisation of quantum many-body scars in the regime $U=\Delta$ of the tilted FH model. We have identified product states $\HG$, $\HGG$ at filling factor $\nu=1$ which give rise to scarred dynamics and reveal towers of ergodicity-breaking many-body eigenstates, allowing to investigate the interplay of many-body scarring with other facets of weak ergodicity breaking such as localisation and Hilbert space fragmentation.  In addition to the filling factor $\nu=1$, we have also studied the filling $\nu=1/2$ used in Ref.~\onlinecite{Scherg}.  In the latter case, taking the large-tilt limit $\Delta\gg U, J$ and using a Schrieffer-Wolff transformation up to third order, we found analogous signatures of scars~\cite{SOM}, provided we neglect the diagonal terms in the effective Hamiltonian. Under these assumptions,  the resulting model can be viewed as a spinful generalisation of the fractional quantum Hall effect on a thin torus~\cite{Moudgalya2019}.  By contrast, the approach presented here for $\nu=1$ is considerably simpler as it allows to conveniently eliminate the undesirable diagonal terms. 

{\sl Acknowledgements.} We acknowledge support by the Leverhulme Trust Research Leadership Award RL-2019-015 (ZP, AH),
 and by EPSRC grants EP/R020612/1 (ZP),  EP/R513258/1 (JYD), and EP/M50807X/1 (CJT).  
  Statement of compliance with EPSRC policy framework on research data: This publication is theoretical work that does not require supporting research data. 
 A.H. acknowledges funding provided by the Institute of Physics Belgrade, through the grant by the Ministry of Education, Science, and Technological Development of the Republic of Serbia. Part of the numerical simulations were performed on the PARADOX-IV supercomputing facility at the Scientific Computing Laboratory, National Center of Excellence for the Study of Complex Systems, Institute of Physics Belgrade.

\bibliography{references}

\onecolumngrid 
\newpage

\begin{center}
{\bf \large Supplementary online material for ``A proposal for realising quantum scars in the tilted 1D Fermi-Hubbard model"}
\end{center}
\begin{center}
Jean-Yves Desaules$^1$, Ana Hudomal$^{1,2}$, Christopher J. Turner$^1$, and Zlatko Papi\'c$^1$\\
\vspace*{0.1cm}
{\footnotesize
$^1$School of Physics and Astronomy, University of Leeds, Leeds LS2 9JT, United Kingdom\\
$^2$Institute  of  Physics  Belgrade,  University  of  Belgrade,  11080  Belgrade,  Serbia
}
\end{center}
\setcounter{subsection}{0}
\setcounter{equation}{0}
\setcounter{figure}{0}
\renewcommand{\theequation}{S\arabic{equation}}
\renewcommand{\thefigure}{S\arabic{figure}}
\renewcommand{\thesubsection}{S\arabic{subsection}}

{\footnotesize In this Supplementary Material, we discuss details of the numerical implementation of the tilted 1D Fermi-Hubbard model using the Jordan-Wigner mapping to spins, the symmetries of the model, and we present more details on the level statistics and  eigenstate properties of the model. We provide further evidence for the importance of the hypergrid subgraph for many-body scarring, including a perturbation which improves the revivals. Finally, we also explain how our results relate to a different regime of the model at high tilt ($\Delta \gg U, J$) and electronic filling $\nu=1/2$.
}

\vspace*{1cm}

\twocolumngrid

\section{Jordan-Wigner transformation and symmetries}\label{app:sym}

\subsection{Jordan-Wigner transformation}

In order to simplify the computations, we perform a Jordan-Wigner transformation to express the Hamiltonian in terms of spin operators. As a convention, we will set that the spin-down fermions are located ``on the right" of the up-spin ones. For open boundary conditions the resulting Hamiltonian for the effective model is :
\begin{equation} \label{eq:H_JW}
\begin{aligned}
\hat{H}_{\rm eff,JW}=&-J\sum_{j}  \hat{S}^{+}_{j,\downarrow}\hat{S}^{-}_{j+1,\downarrow}\hat{P}_{j,\uparrow} \left( 1-\hat{P}_{j+1,\uparrow}\right)\\
&+J\sum_{j}  \hat{S}^{+}_{j,\uparrow}\hat{S}^{-}_{j+1,\uparrow}\hat{P}_{j,\downarrow} \left(1-\hat{P}_{j+1,\downarrow}\right)
\end{aligned}
\end{equation}
where 
\begin{equation*}
\hat{S}^z=\begin{pmatrix}1 & 0 \\ 0 & -1 \end{pmatrix}, \quad \hat{S}^+=\begin{pmatrix}0 & 1 \\ 0 & 0 \end{pmatrix} \quad {\rm and} \quad  \hat{S}^-=\begin{pmatrix}0 & 0 \\ 1 & 0 \end{pmatrix}
\end{equation*}
are the usual Pauli spin operators and $\hat{P}_{j,\sigma}=\frac{1+\hat{S}_{j,\sigma}^z}{2}$ is the projector on the excited state of the Jordan-Wigner spin.
In order to minimise the confusion between the spin of the original fermions and the Jordan-Wigner spins we will use ``spin" to only refer to the fermion spin. For the Jordan-Wigner spins we will describe the physics in term of ``excitations", e.g.,  the action of $\hat{S}^+_{j,\downarrow}$ is to create an excitation on site $j$ with spin down.

\subsection{Symmetries}

In the component of the Hilbert space we are interested in (i.e., the one containing the state with alternating $\uparrow$ and $\downarrow$ fermions, as explained in the main text), the Hamiltonian in the Jordan-Wigner spin representation has two $\mathbb{Z}_2$ symmetries. The first one, denoted as $\hat{z}_1$, is given by the joint action of the doublon parity operator and the spin inversion operator: 
\begin{equation}
	\hat{z}_1=\prod_{j=1}^{N} (-1)^{\hat{P}_{j,\downarrow}\hat{P}_{j,\uparrow}} \left(\hat{S}_{j,\uparrow}^+ \hat{S}_{j,\downarrow}^- {+}\hat{S}_{j,\downarrow}^+ \hat{S}_{j,\uparrow}^- {+}\frac{1{+}\hat{S}^z_{j,\uparrow}\hat{S}^z_{j,\downarrow}}{2} \right)
\end{equation}
The doublon parity is diagonal in the product basis and gives +1 if there is an even number of doublons and -1 otherwise. The spin inversion simply changes the excitations between up spin and down spin. 

The second symmetry, denoted as $\hat{z}_2$, corresponds to the joint application of the spatial inversion operator and of the particle-hole operator: 
\begin{equation}
\begin{aligned}
\hat{z}_2=\prod_{j=1}^{N/2}\prod_{\sigma}\Big(&\hat{S}_{j,\sigma}^+ \hat{S}_{N-j-1,\sigma}^+ +\hat{S}_{j,\sigma}^- \hat{S}_{N-j-1,\sigma}^-  \\
&+\frac{1-\hat{S}^z_{j,\sigma}\hat{S}^z_{N-j-1,\sigma}}{2}\Big).
\end{aligned}
\end{equation}
The spatial inversion swaps sites $j$ and $N-j$, while the particle-hole conjugation puts an excitation in every empty site and vice-versa. 
Performing a Jordan-Wigner transformation on the full Fermi-Hubbard model with the same fermion ordering convention produces a spin Hamiltonian which has the symmetry generated by $\hat{z}_1$ but not by $\hat{z}_2$.

Working in the fermionic language, the corresponding set of symmetry operators are
\begin{equation}
  \hat{y}_1= \prod_j e^{-i\pi \frac{1}{2}(i \hat{c}^\dagger_{j,\downarrow} \hat{c}_{j,\uparrow} - i \hat{c}^\dagger_{j,\uparrow} \hat{c}_{j,\downarrow})}
\end{equation}
and
\begin{equation}
  \hat{y}_2 = \prod_j  (-1)^{\hat{n}_{j,\uparrow}\hat{n}_{j,\downarrow}}  \mathcal{P} \mathcal{R}\text{,}
\end{equation}
using the particle-hole operator $\mathcal{P}$ and spatial-reflection operator $\mathcal{R}$ defined by the adjoint actions,
\begin{align}
  \mathcal{P} c_{j,\sigma}^\dagger &= c_{j,\sigma} \mathcal{P}\text{,} &
  \mathcal{R} c_{j,\sigma}^\dagger &= c_{N-j-1,\sigma}^\dagger \mathcal{R}\text{.}
\end{align}
The particle-hole operator also has a non-trivial action on the vacuum state
\begin{equation}
  \mathcal{P} \ket{0} = \prod_j c_{j,\uparrow}^\dagger c_{j,\downarrow}^\dagger \ket{0}\text{.}
\end{equation}
These are equivalent to $\hat{z}_1$ and $\hat{z}_2$ respectively, up to an overall spin rotation and some charge-dependent phase factors.
The first one is similar to the spin-reversal symmetry from Eq.~(2.57) in Ref.~\cite{1D_Hubbard}.

\subsection{Negative matrix elements}

The spin formulation of the model, after the Jordan-Wigner transformation, offers a convenient framework for numerics. However, it is clear that the Hamiltonian in Eq.~(\ref{eq:H_JW}), expressed in the product state basis, has both negative as well as positive matrix elements. In order to treat this model as an unweighted graph, we show that a simple basis transformation is enough to get rid of all minus signs in the Hamiltonian. These occur only when moving a spin down excitation. Here we propose the following convention for fixing the signs: we assign to each product a sign equal to the parity of the number of spin down excitations on the even sites. This corresponds to a change of basis using the diagonal matrix $\hat{T}=\sum_{j=1}^{N/2}(-1)^{\hat{P}_{j,\downarrow}}$. If a spin up excitation is moved, this does not change this quantity and it follows that both states connected by this move have the same sign. Whether it is +1 or -1 is irrelevant as the matrix element will get a factor equivalent to their product, which is always +1. On the other hand, if a spin down excitation is moved, this parity number will always change by 1, and the states connected by the move will have opposite signs. From this, it follows that all the matrix elements of 
$\hat{T}\hat{H}_{\rm eff,JW} \hat{T}$ are equal to +1, thus we can view the model in terms of an undirected, unweighted graph.

Alternatively, the same result can be obtained by choosing a different convention for the Jordan-Wigner transformation.
Instead of interleaving the fermions with a different spin-projection, one can first place all up-spin fermions and then all down-spin ones with a simple linear order for each species.
This has the effect of ensuring that each of the `hopping' terms never takes a fermion past another.
Then, all these matrix elements are equal and can be brought to $1$ by the choice of the coupling constant $J$.

\begin{figure}[thb]
	\centering
	\includegraphics[width=\linewidth]{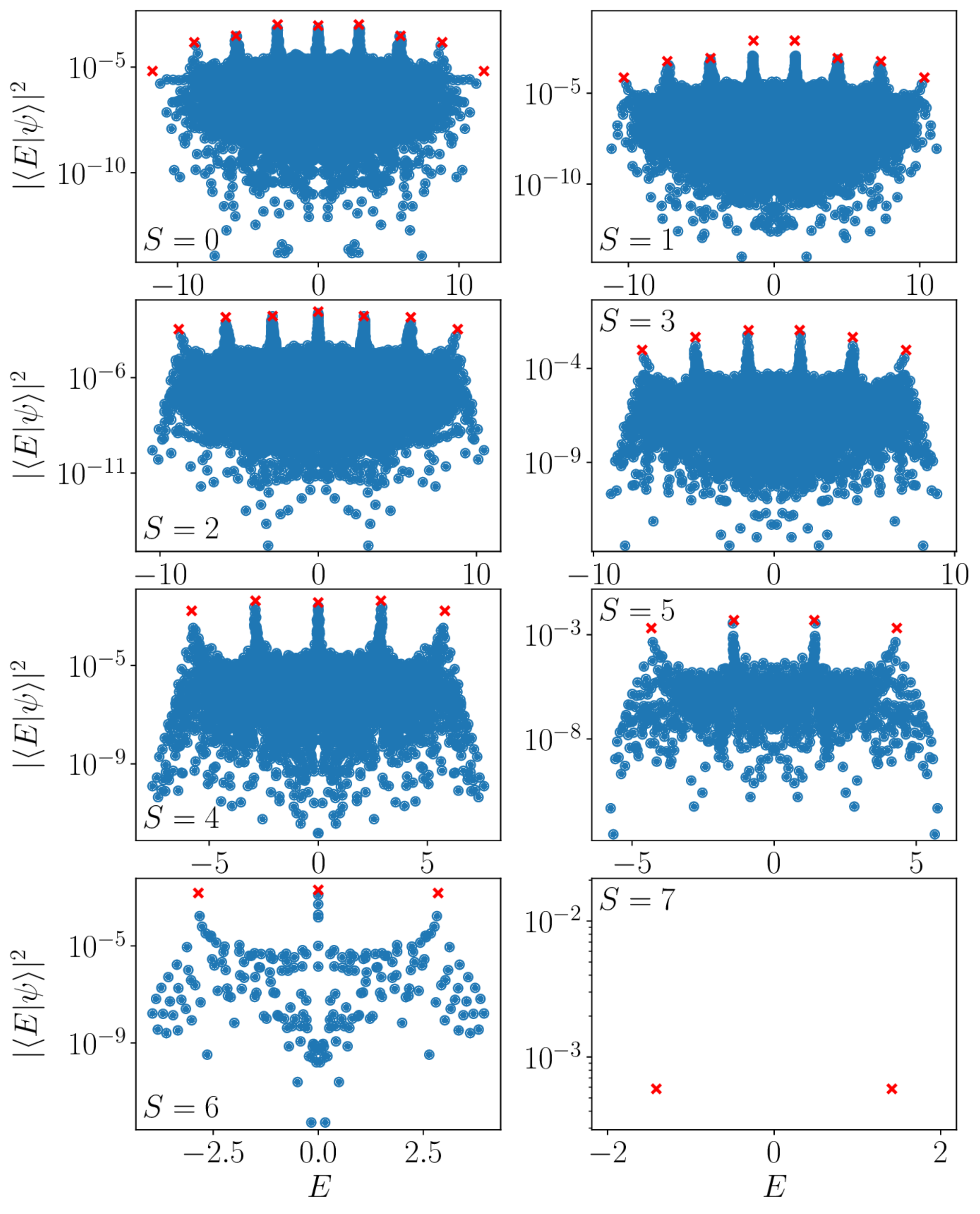}
	\caption{\small Overlap of the eigenstates with  $\HG$ state in the effective model for $N=16$ in each total spin sector. The red crosses indicate the states at the top of each tower in the relevant sector.  
	}
	\label{fig:olap_sects}
\end{figure}

\begin{figure}[thb]
	\centering
	\includegraphics[width=\linewidth]{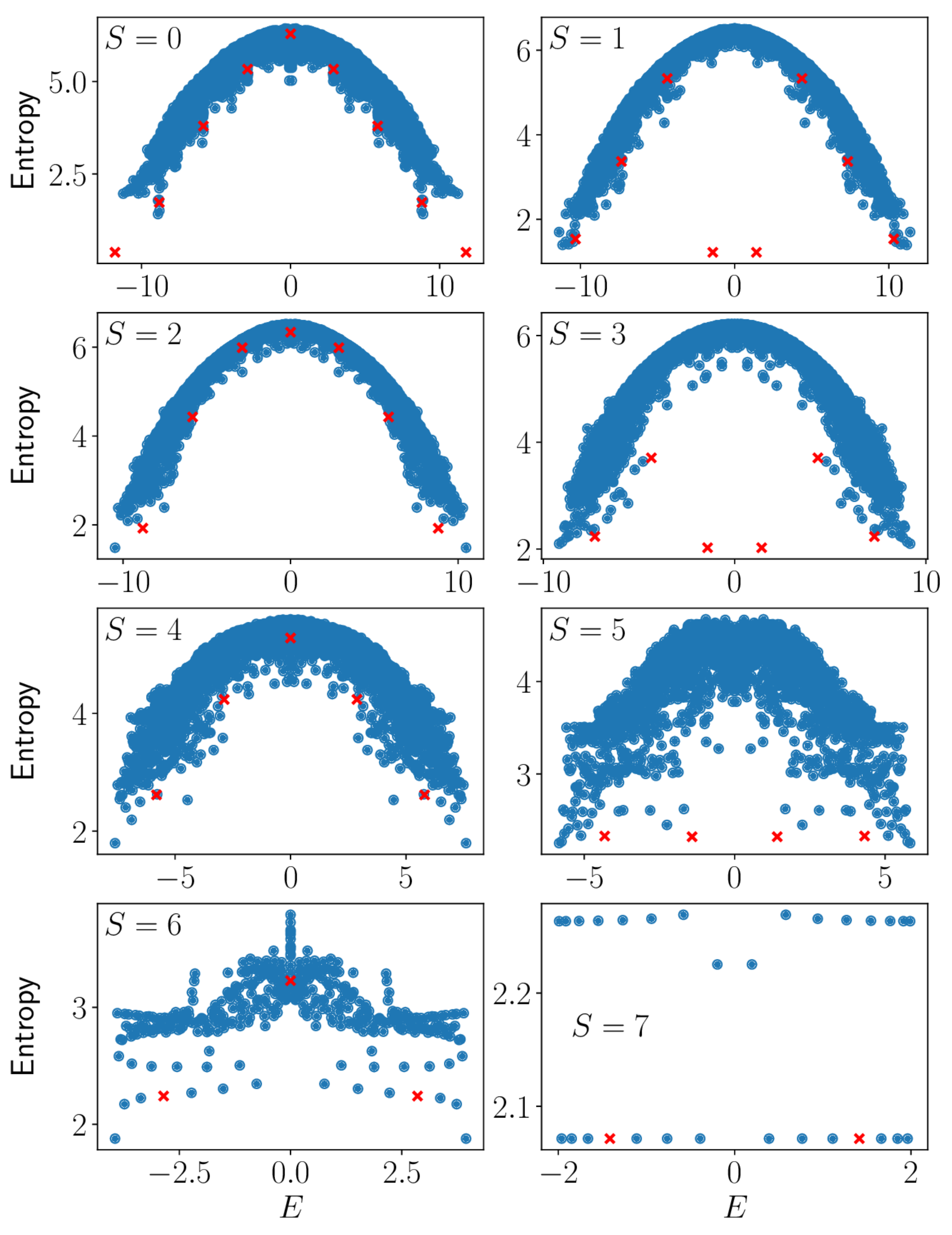}
	\caption{\small Entropy of the eigenstates with  $\HG$ state in the effective model for $N=16$. The red crosses indicate the states at the top of each tower in the relevant sector and are the same as in \Cref{fig:olap_sects}.
	}
	\label{fig:Sent_sects}
\end{figure}

\section{Individual symmetry sectors}\label{sec:level_stat}

Because of SU(2) symmetry, the number of total spin sectors is extensive in system size. In Fig.~\ref{fig:olap_sects} we show results for the overlap of eigenstates with $\HG$ state, while Fig.~\ref{fig:Sent_sects} shows the entropy of eigenstates in different total spin sectors. We focus on system size $N=16$ and neglect the sector with maximum total spin ($S=8$), since this only has a single state. 
All sectors display towers of states  with the same approximate spacing of energy equal to $2\sqrt{2}$ (\Cref{fig:olap_sects}). However as the total spin increases the number of towers decreases as $N/2+1-S$. Because of the change in parity of that quantity, the location in energy of the towers is also dependent on the parity of the total spin. Sectors with an odd number of towers will have them at $E\approx \pm n 2\sqrt{2}$, with $n$ being a positive integer or zero. On the other hand, sectors with an even number of towers will have them at $E\approx \pm(1+2n)\sqrt{2}$.

For the largest sectors ($S\leq 4$) the entropy band is relatively narrow, with only a few outliers corresponding to the top state of some of the towers (\Cref{fig:Sent_sects}). Interestingly, the states with a very low entropy are only present for odd total spin for $N=16$. This is linked with the dependence of the location on the towers on the total spin. Indeed, the most atypical states have a fixed energy $E\approx \pm \sqrt{2}$ for all system sizes attainable by exact diagonalisation. Hence they will only be present in sectors with an even number of towers, which means an odd total spin for $N=16$.

\begin{figure}[thb]
\centering
\includegraphics[width=\linewidth]{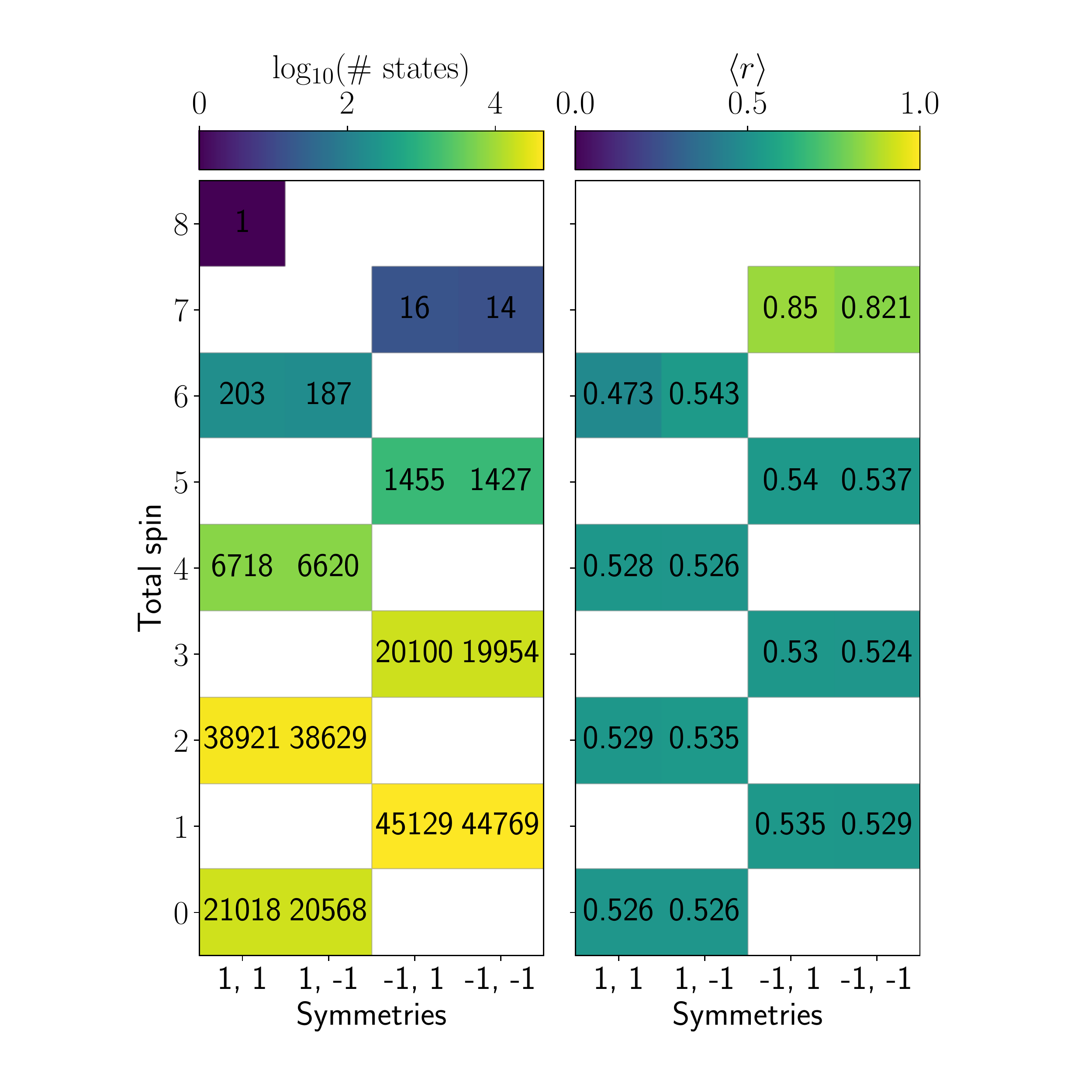}
\caption{\small Number of states and level statistics in each symmetry sector $(z1,z2)$ in the effective model for $N=16$.
}
\label{fig:app_all_r}
\end{figure}

\begin{figure}[thb]
\centering
\includegraphics[width=0.9\linewidth]{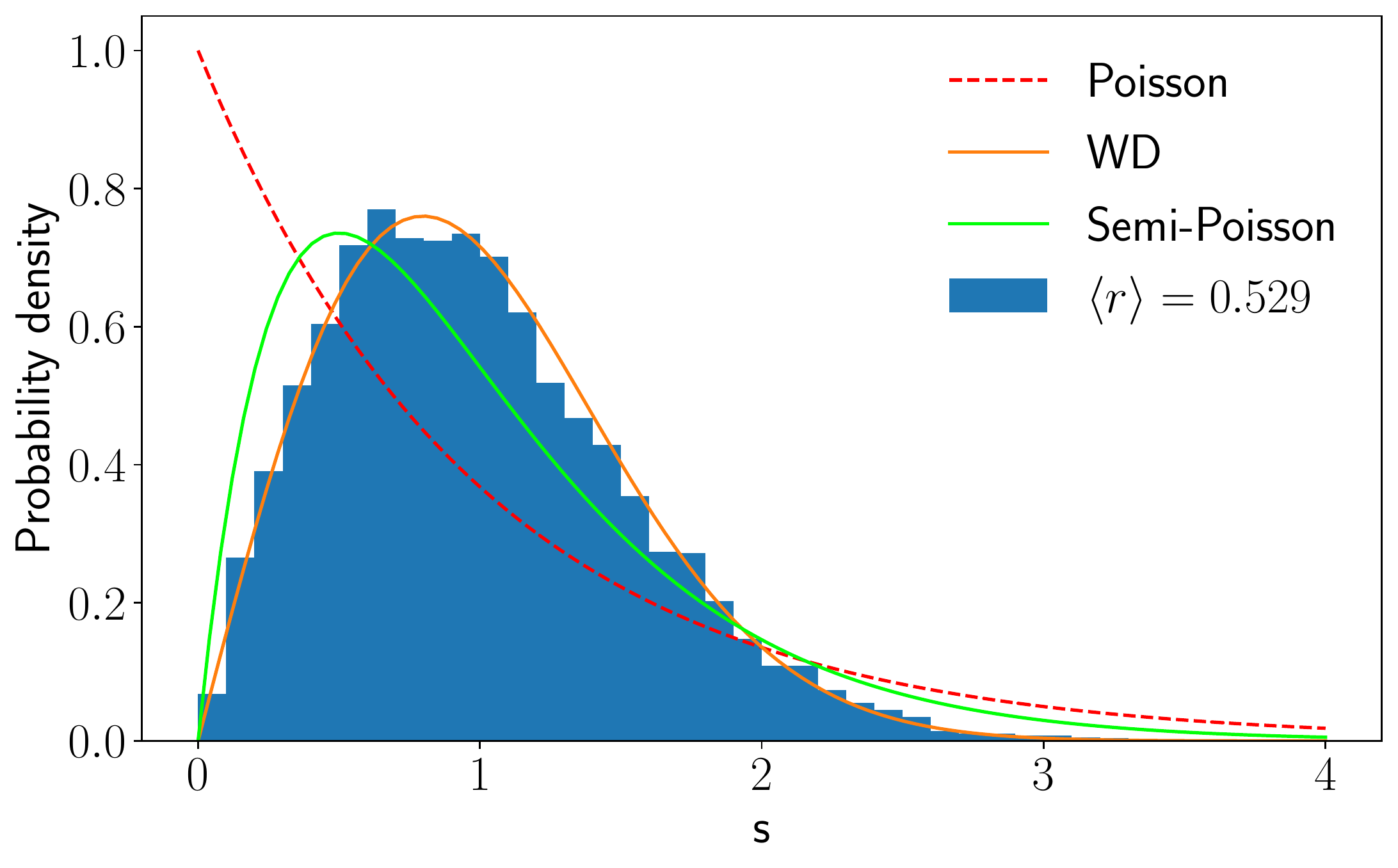}
\caption{\small Distribution of the energy level spacing after unfolding \cite{Mehta2004}  in the sector $z_1=1$, $z_2=1$, $S=2$ for $N=16$. The result for this sector is very close to a Wigner-Dyson distribution.
}
\label{fig:level_stat}
\end{figure}

After all the symmetries have been resolved, one can study energy level statistics in each symmetry sector separately (\Cref{fig:app_all_r}). 
All sectors with a large number of states ($\mathcal{D}>10^3$) have mean level spacing $\langle r \rangle$ very close to 0.53, compatible with the Wigner-Dyson distribution. 
We also show a histogram of level spacing after unfolding \cite{Mehta2004}. This is done for the sector $z_1=1$, $z_2=1$, $S=2$ in Figure~\ref{fig:level_stat}. This sector was chosen because it has the largest overlap with $\HG$ state at the system size studied and also has a large number of states. The histogram of the level spacing confirms that this sector indeed corresponds to a chaotic quantum system.

\section{Influence of the hypergrid}

The influence of the hypergrid subgraph is visible in the dynamics, i.e., in the revivals from a few product states as demonstrated in the main text. In this section we illustrate the impact of the hypergrid on eigenstates and the long-time behaviour of the system. 

\begin{figure}[thb]
\centering
\includegraphics[width=0.9\linewidth]{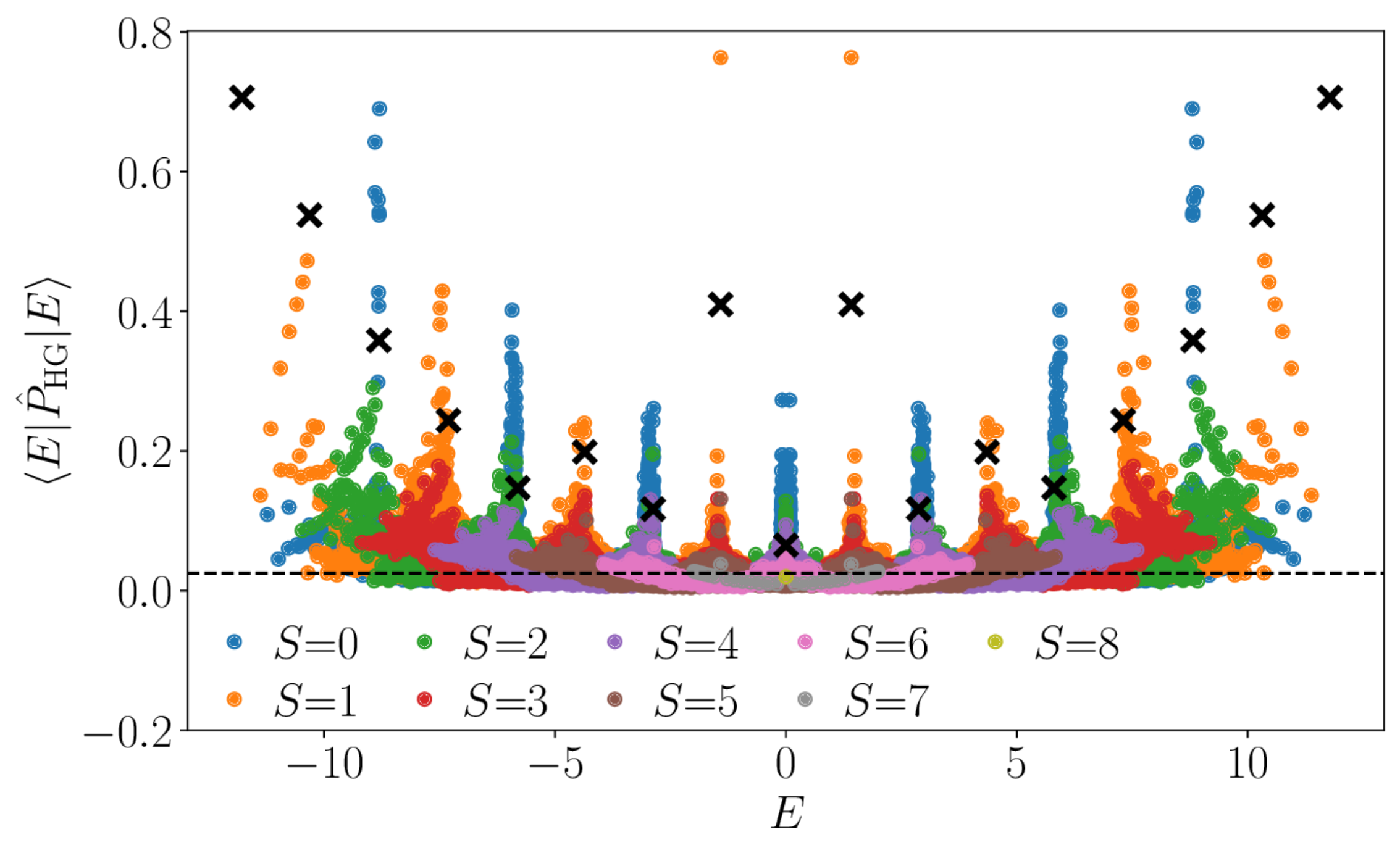}
\caption{\small Expectation value of the projector onto the hypergrid for the eigenstates at $N=16$. The colours indicate the value of total spin for each eigenstate and the dashed line is the average value of $P_{\rm HG}$ for the system.
Crosses are those states with highest overlap around their energy irrespective of spin.}
\label{fig:proj_eigs}
\end{figure}

In a fully ergodic system, one expects the support of an eigenstate on a subset of states to be approximately equal to the ratio of the number of states belonging to this subset and  the total dimension of the Hilbert space. Performing this computation for the effective model and choosing the hypergrid subgraph as the subset shows anomalous concentration of some eigenstates -- see~\Cref{fig:proj_eigs}. 
As the overlap with the $\HG$ state forms a lower bound to the support on the hypergrid, we find the same kind of towers of states, located at the same energies, in this plot. However, where the maximum overlap with $\HG$ was approximately 0.011 for $N=16$, the maximum support on the hypergrid is instead close to 0.76 for the same size. This confirms that the hypergrid subgraph still leaves a large imprint on the eigenstates, even when it only comprises less than $2.5\%$ of the Hilbert space for $N=16$.

\begin{figure}[thb]
\centering
\includegraphics[width=0.9\linewidth]{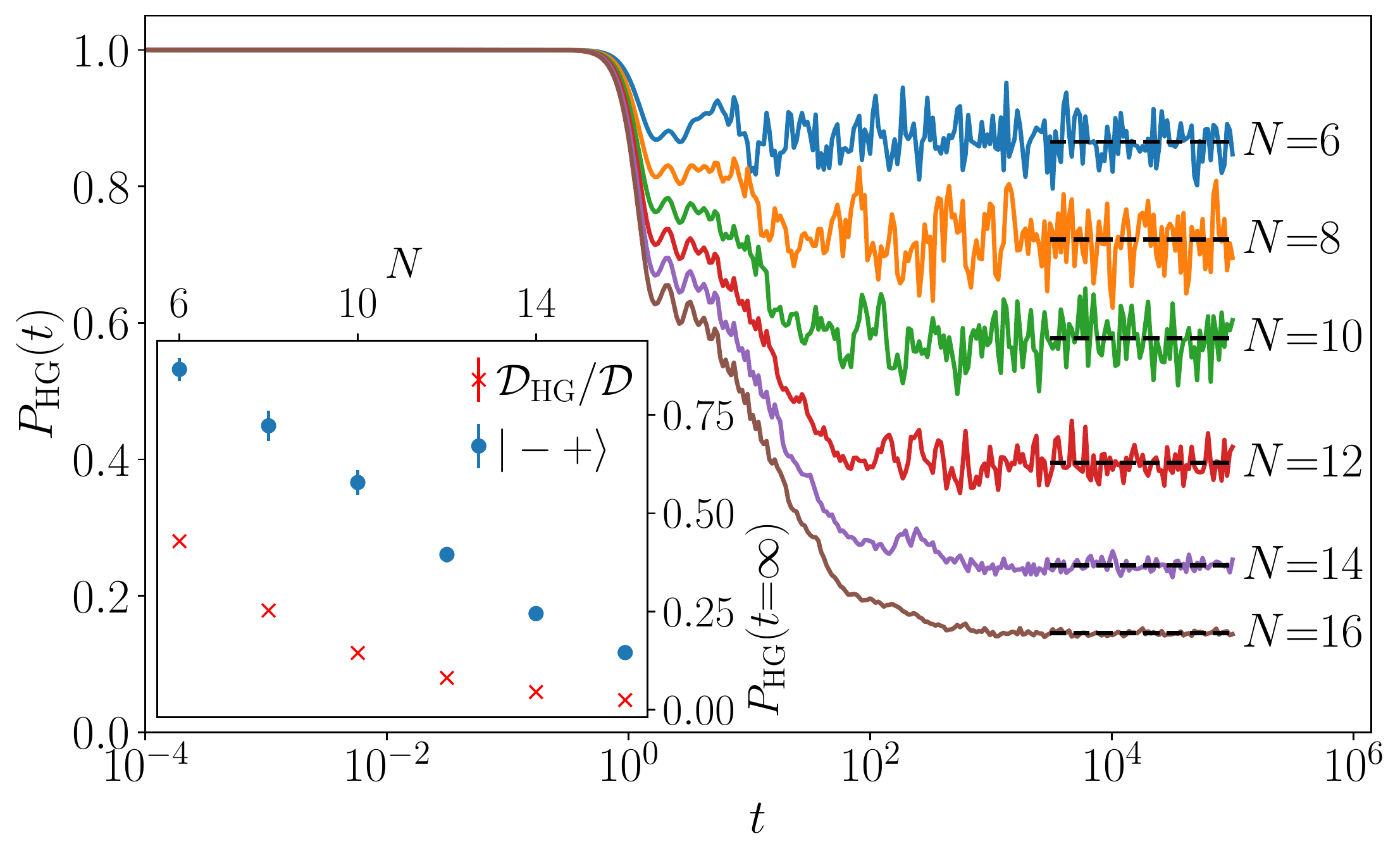}
\caption{\small Expectation value of the projector on the hypergrid states over time for various system sizes with the initial state $\HG$. The dotted black lines show the average of this projector at long times. The inset shows these long-time value compared to the expected value from the relative size of the hypergrid for various system sizes.}
\label{fig:proj_evol}
\end{figure}

\begin{figure}[thb]
\centering
\includegraphics[width=0.9\linewidth]{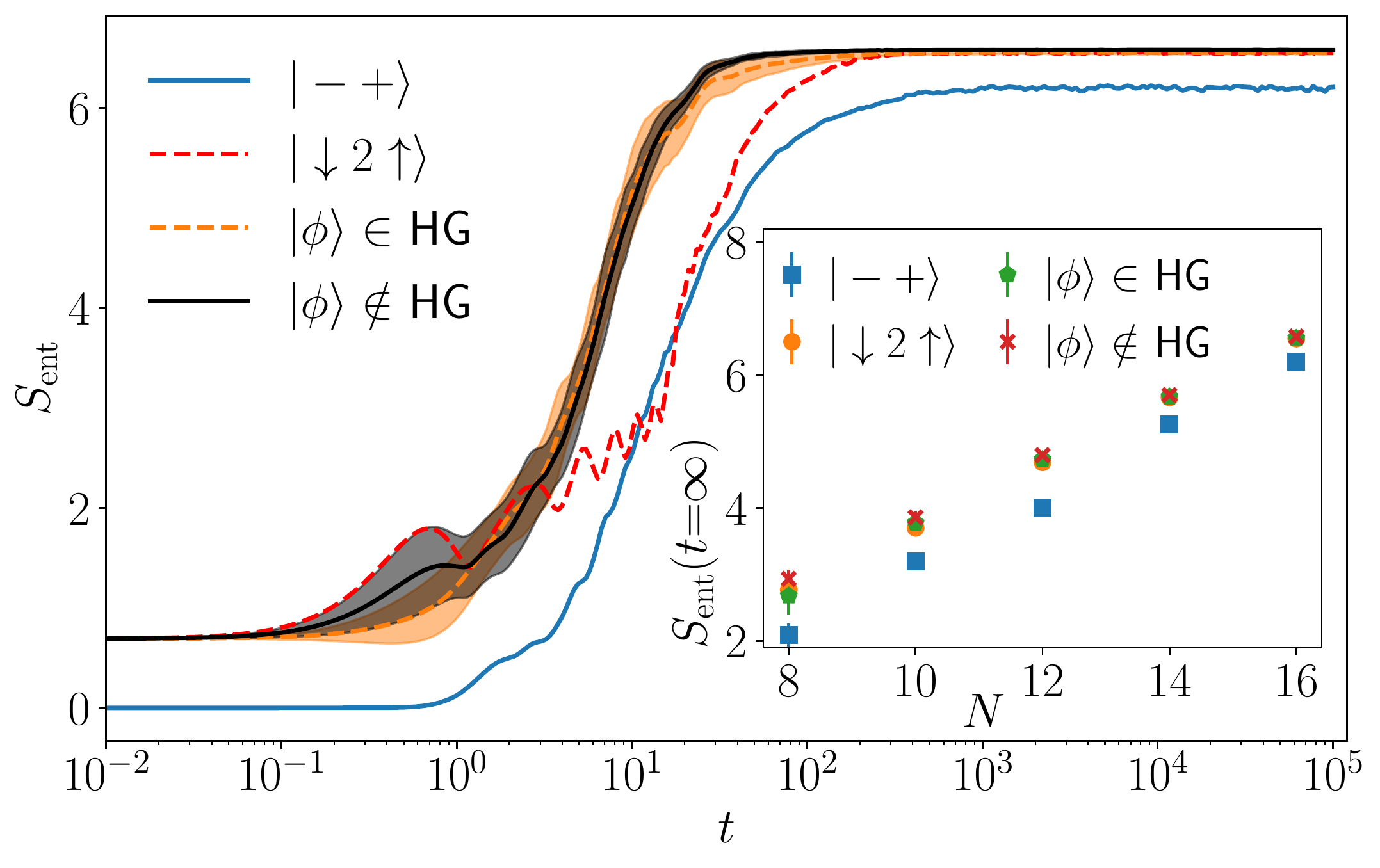}
\caption{\small Entanglement entropy as a function of time for different initial states and system size $N=16$.  $\ket{\phi}\notin$ HG and $\ket{\phi}\in$ HG denote the average over 10 random product states, respectively outside of and within the hypergrid, that have been projected into the two symmetry sectors in which the $\HG$ state has support. This was done to remove the influence of these symmetries in the entanglement entropy. The shaded area corresponds to the standard deviation. The inset shows the long-time value for different initial states as a function of system size. For the system sizes investigated, all states except $\HG$ converge towards the same entropy value.}
\label{fig:S_evol_log}
\end{figure}

Another signature of non-ergodicity caused by the hypergrid subgraph can be seen in the long-time expectation value of the projector onto the hypergrid, $P_{\mathrm{HG}}(t)$, when starting from the $\HG$ state. It converges to a non-zero value which is higher than what would be expected from the relative size of the hypergrid in the Hilbert space (\Cref{fig:proj_evol}).
This hints that the embedding of the hypergrid in the effective model is non-trivial and that the long-lived oscillations in the dynamics cannot simply be explained by the relative size of this substructure.
This can also be seen in the long-time value of the entanglement entropy (\Cref{fig:S_evol_log}). For a large enough system, most states saturate to the same entropy but for $\HG$ this long-time value is lower, showing that the wavefunction is not entirely spread into the whole Hilbert space. As is demonstrated in the inset of \Cref{fig:S_evol_log}, this lower saturation entropy is not visible for generic states within the hypergrid. It is also interesting to note that the long-time entropy value of $\HG$ is still higher than what would be expected from a system with the same dimension as the hypergrid.

\section{Full model}

\begin{figure}[htb]
\centering
\includegraphics[width=0.8\linewidth]{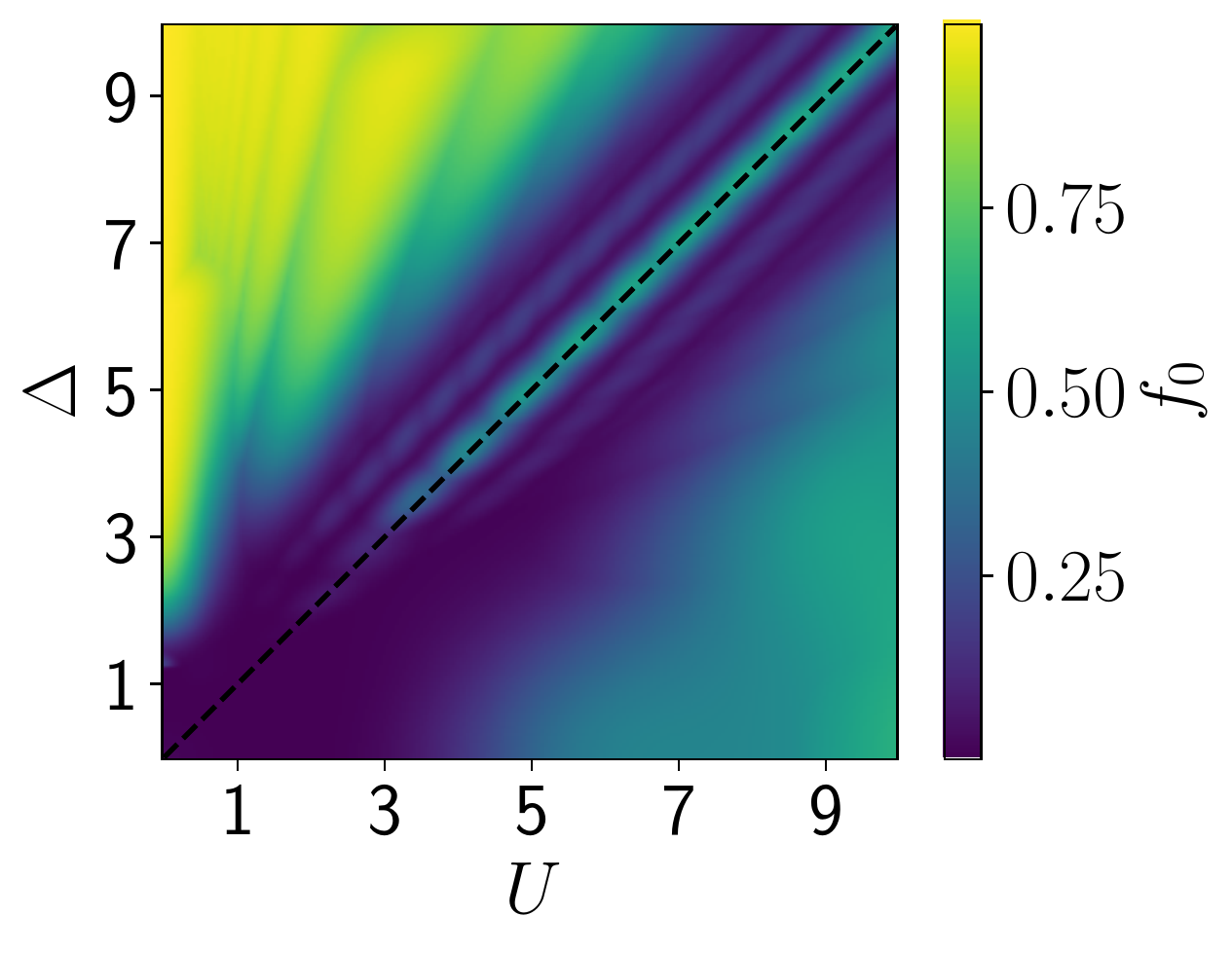}
\caption{\small Maximum revival fidelity in the full model between $t=1$ and $t=10$ when starting from the $\HG$ state. The revivals show an increasingly better fidelity on the $U=\Delta$ diagonal as the value of these parameters get larger. The high fidelity in the top left corner corresponds to the high-tilt regime investigated in \cite{Scherg}, where a large number of states show revivals due to the conservation of the dipole moment and of the number of doublons \emph{separately}.}
\label{fig:fid_heatmap}
\end{figure}

In this section we provide further results showing that the scarred dynamics persists in the full model in Eq.~(1) of the main text at finite values of $\Delta$ and $U$.
In our analysis of the effective model, we looked at the fidelity of wavefunction revivals; here we show that the wavefunction revivals are also visible in the full model for a relatively broad range of parameters $U$ and $\Delta$ (\Cref{fig:fid_heatmap}). 

\begin{figure}[htb]
	\centering
	\includegraphics[width=\linewidth]{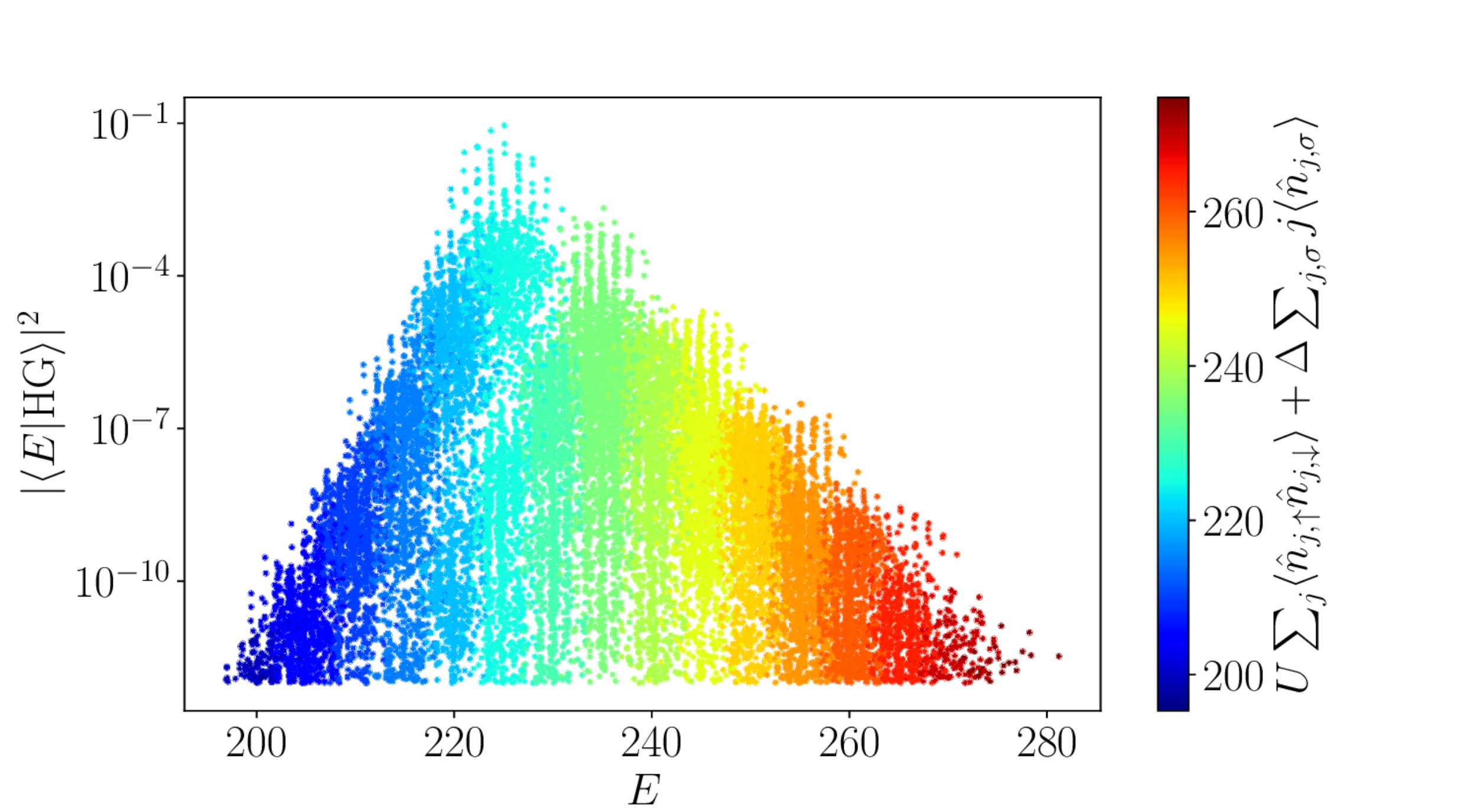}
	\caption{\small Overlap of  $\HG$ state with eigenstates of the full model for $N=10$, $U=\Delta=5$, and $J=1$. The colour indicates the expectation value of the number of doublons plus the dipole moment. We see that already at this value of the parameters the eigenstates start to form distinct towers.}
	\label{fig:olap_full}
\end{figure}
We also plot the overlap of $\HG$ state with the eigenstates of the full model in \Cref{fig:olap_full}.
Colouring each eigenstate with the expectation value of $U\sum_j\langle\hat{n}_{j,\uparrow}\hat{n}_{j,\downarrow}\rangle+\Delta\sum_{j,\sigma}j\langle\hat{n}_{j,\sigma} \rangle $ highlights the fact that the eigenstates start to form distinct towers at these parameter values. Each of these towers correspond to an expectation value of this operator with a value close to an integer multiplied by $U=\Delta$. As this parameter is increased the mixing between the towers is reduced until they are completely separated in the $U=\Delta=\infty$ limit. In this case only eigenstates in the central tower have a non-zero overlap with the $\HG$ state.

\section{Improving the revival by perturbation}

\begin{figure}[thb]
	\centering
	\includegraphics[width=0.9\linewidth]{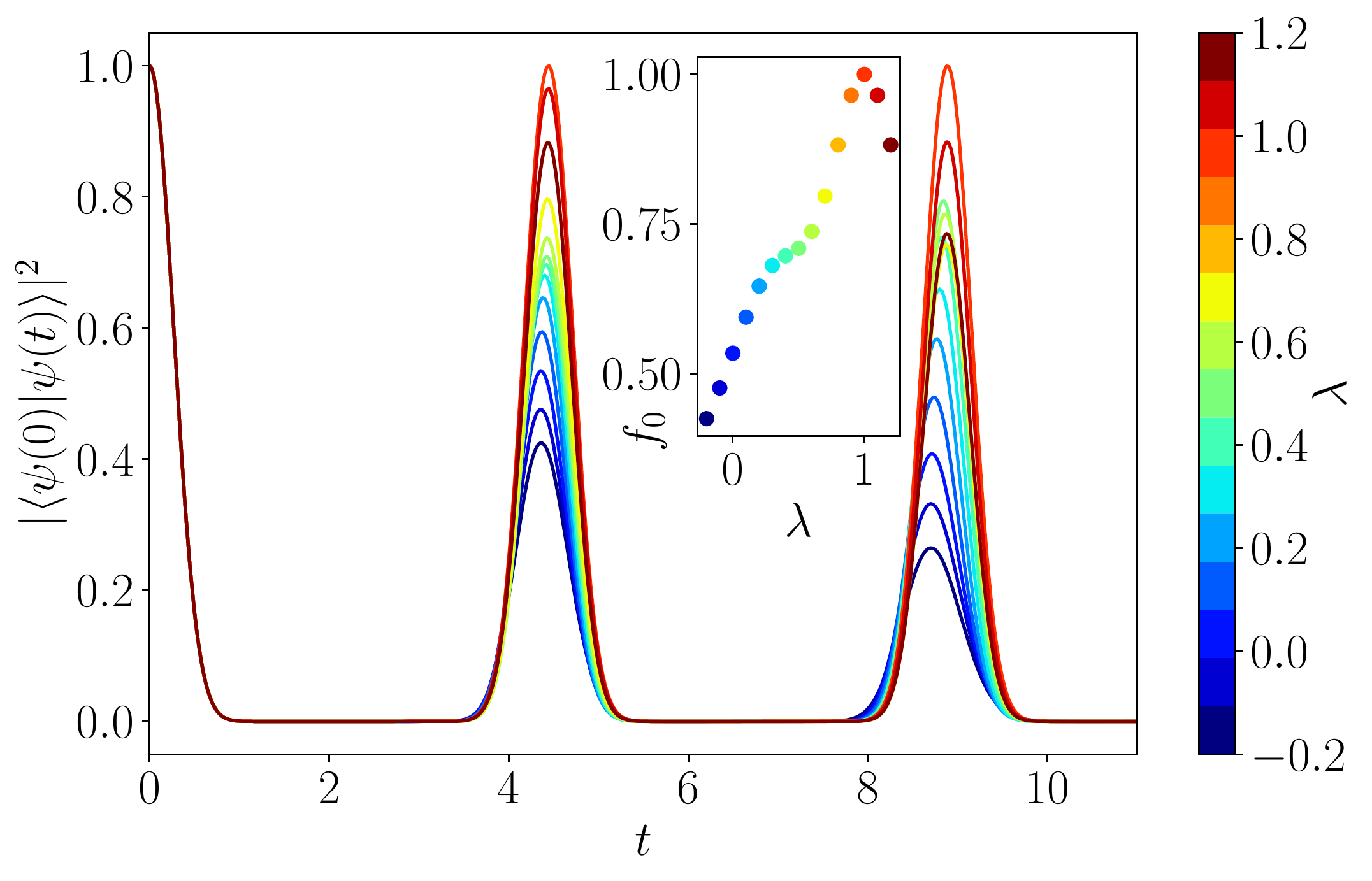}
	\caption{\small Fidelity revivals for the perturbed effective model for $N=14$. At $\lambda=1$ the hypergrid subgraph is completely isolated from the rest of the Hilbert space and the revivals from the $\HG$ state (and from any other of the hypergrid corners) become perfect.}
	\label{fig:pert_revs}
\end{figure}

As the revivals emerge due to an embedded subgraph, we can devise a perturbation that enhances the revivals by decreasing the leakage out of the hypergrid. For the effective model, this can be achieved by a dimerised  perturbation given by 
\begin{equation}
	H_{\rm Pert, eff}{=}\lambda J\hspace{-0.24cm}\sum_{j=1}^{N/2-1}\hspace{-0.1cm} \sum_{\sigma} \hat{c}^{\dagger}_{2j,\sigma}\hat{c}_{2j+1,\sigma}\hat{n}_{2j,\overline{\sigma}}(1-\hat{n}_{2j+1,\overline{\sigma}}) {+} \mathrm{h.c.},
\end{equation}
i.e., the perturbation only affects interaction between the unit cells, leaving the hypergrid subgraph untouched.  For $\lambda=1$, the hopping term between cells in the Hamiltonian and in the perturbation cancel each other. As a result, the hypergrid  becomes completely isolated from the rest of the Hilbert space, resulting in the perfect wavefunction revival. This can be seen in Figure~\ref{fig:pert_revs} for $N=14$.
These results show clearly that the perturbation has a very weak effect on the period of the revivals, but only modulates their amplitude. Furthermore, as expected the fidelity peaks at $\lambda=1$, when the system revives perfectly.

The same perturbation can also be applied to the full model, in which case it takes the form
\begin{equation}
	\hat{H}=\lambda J\sum_{j=1}^{N/2-1}\sum_{\sigma}\hat{c}^\dagger_{2j,\sigma}\hat{c}_{2j+1,\sigma}+{\rm h.c.}.
\end{equation}
The optimal value of $\lambda$ for the first peak of the fidelity revivals is also present at $\lambda=1$, but the revivals are no longer perfect (the first peak reaches $\sim 70\%$). Furthermore, unlike in the effective model, looking at the second revival peak gives an optimal value of lambda closer to 0.3. 
\begin{figure}[thb]
	\centering
	\includegraphics[width=\linewidth]{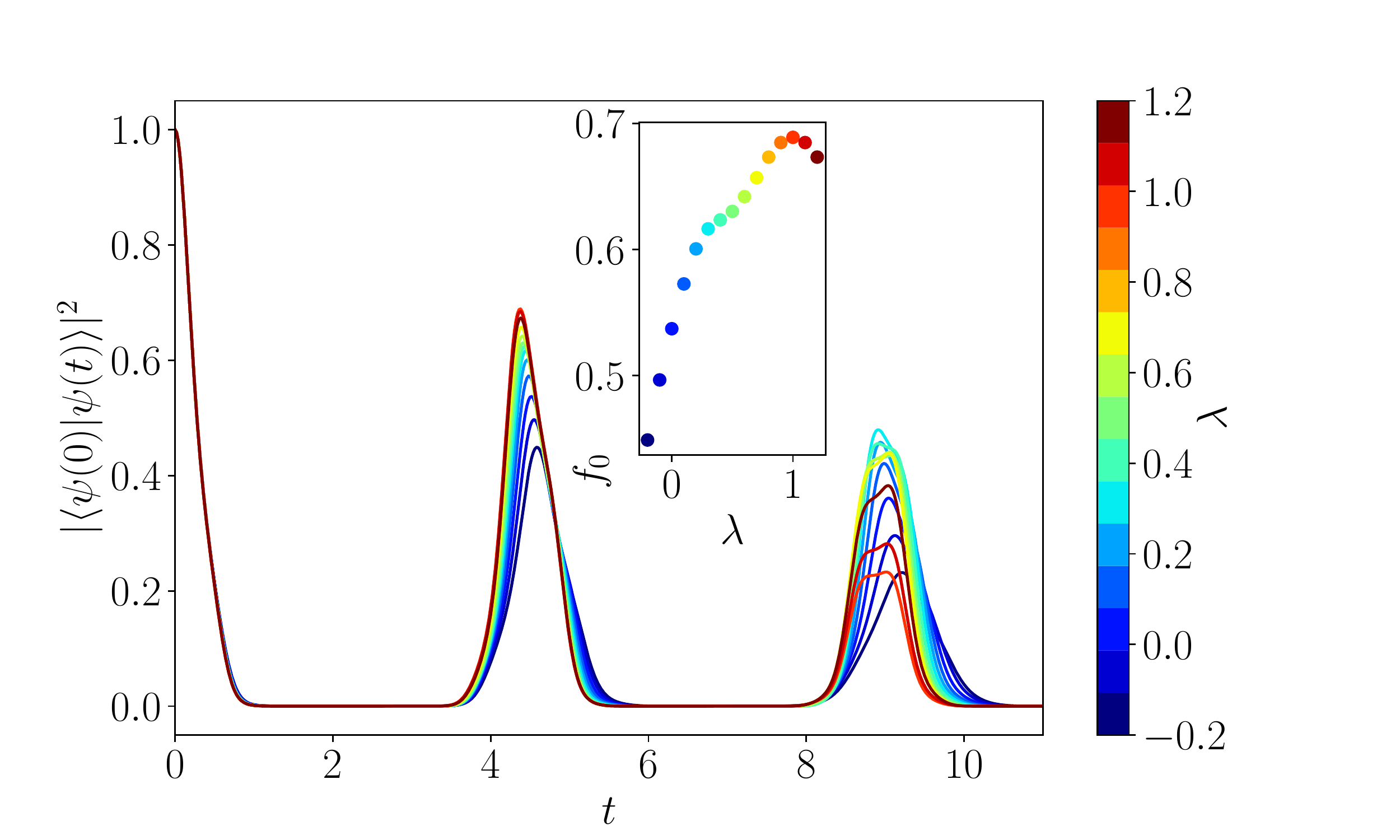}
	\caption{\small Fidelity revivals for the full model with perturbation for $N=12$ starting from the $\HG$ state. As for the effective model $\lambda=1$ gives the best revivals for the this state.}
	\label{fig:pert_revs_full}
\end{figure}
The fact that this perturbation still significantly improves the revivals in the full model strongly suggests that the oscillatory is also due to the imprint of the hypergrid in this case.

\section{Filling $\nu=1/2$}

While most of our work has focused on the sector with filling $\nu=1$, we find similar dynamics in the $\nu=1/2$ sector which was explored in Ref.~\cite{Scherg}. In the high tilt regime $\Delta\gg U,J$,  the effective Hamiltonian is given by
\begin{align} 
	\label{eq:H3_dipole} 
	\hat H_{\textrm{eff}}^{\mathrm{dip}}= &J^{(3)} \hat T_3
	  +U\Big( 1 -\frac{4 J^2}{\Delta^2}\Big)\sum_{i} \hat n_{j,\uparrow} \hat n_{j,\downarrow} \nonumber \\ 
	&+ 2 J^{(3)} \hat T_{XY} + 2 J^{(3)} \sum_{j,\sigma} \hat n_{j,\sigma} \hat n_{j+1,\bar{\sigma}},
\end{align}
up to some constant terms, where
\begin{align} \label{H3exp}
    &J^{(3)}=\frac{J^2U}{\Delta^2} \\
	&\hat T_3= \sum_{i,\sigma}{ \hat c_{i,\sigma} \hat c_{i+1,\sigma} ^{\dagger} \hat c_{i+1,\bar{\sigma}} ^{\dagger} \hat c_{i+2,\bar{\sigma}}} +\text{h.c.},\\ 
	&\hat T_{XY} =  \sum_{i,\sigma} \hat c_{i,\bar{\sigma}}^{\dagger} \hat c_{i+1,\bar{\sigma}} \hat c_{i+1,\sigma}^{\dagger}  \hat c_{i,{\sigma}} 
\end{align}
The action of this Hamiltonian also produces a  fragmentation of the Hilbert space, and we focus on the sector with $N/4$ fermions with spin down and $N/4$ with spin up, and containing the charge density wave (CDW) state studied in experiment with  fermions only on even sites.

The effective Hamiltonian projected to the sector of the above CDW state can be further simplified by noticing that some terms exactly vanish. The $\hat{T}_3$ operator only allows to squeeze two fermions with opposite spin to create a doublon between them. As all fermions are originally separated by an empty site it is impossible for this operator to reach a state with neighbouring fermions: they will always be either on the same site or with an empty site between them. Hence, the operator $\hat{T}_{XY}$ and the potential terms $\hat{n}_{j,\sigma} \hat n_{j+1,\bar{\sigma}}$ are always zero. As the doublon-counting term is diagonal, the only term that allows for hopping between product states is $\hat{T}_3$. By studying the effect of this operator in a 4-site cell we will show that it leads to the same constrained Hilbert space structure as the sector studied at filling $\nu=1$.

Indeed, for $\nu=1$ with two sites we have the possible states $\ket{\uparrow\downarrow} \leftrightarrow \ket{\updownarrow 0} \leftrightarrow \ket{\downarrow\uparrow}$ forming a three level system.
For $\nu=1/2$, the operator $\hat{T}_3$ also only allows 3 states but now for four sites as $\ket{\uparrow 0 \downarrow 0} \leftrightarrow \ket{0 \updownarrow 0 0} \leftrightarrow \ket{\downarrow{}0\uparrow{}0}$. As all actions of the Hamiltonian can be understood in terms of these cells, it ensues that both models
 have \emph{exactly} the same off-diagonal matrix elements if $N_{\nu=1/2}=2N_{\nu=1}$. The only difference is the diagonal term counting the number of doublons for $\nu=1/2$ filling. 

To better grasp the consequences of this potential term, it is useful to rewrite the effective Hamiltonian in Eq.~(\ref{eq:H3_dipole}) without the redundant terms and factorising the common prefactors to obtain
\begin{equation}
\begin{aligned}\label{eq:H_CDW}
 \hat H_{\textrm{CDW}}&= J^{(3)} \hat T_3
 	  +U\Big( 1 -\frac{4 J^2}{\Delta^2}\Big)\sum_{i} \hat n_{j,\uparrow} \hat n_{j,\downarrow}\\
       &=J^{(3)}\left[\hat T_3
        	  +\Big(\frac{\Delta^2}{J^2} -4\Big)\sum_{i} \hat n_{j,\uparrow} \hat n_{j,\downarrow} \right].
\end{aligned}
\end{equation}
As the Hamiltonian in Eq.~(\ref{eq:H3_dipole}) is only valid for $\Delta \gg U,J$, this is also the case for the one in Eq.~(\ref{eq:H_CDW}). However, it is easy to see that as $\Delta/J$ increases, the potential term will dominate the dynamics. So there is no way to obtain the Hamiltonian $\hat{T}_3$ on its own. Set $\Delta$ too low and the higher order terms in the Schrieffer-Wolff transformation will not be negligible, set it too high and the potential term will dominate the dynamics.
On the other hand, this issue does not arise in the $\nu=1$ sector as only the equivalent of $\hat{T}_3$ is present in the effective Hamiltonian at leading order.

\end{document}